\documentclass[11pt,a4paper]{article}
\usepackage{t1enc}
\usepackage[latin1]{inputenc}
\usepackage[english]{babel}
\usepackage{harvard}
\usepackage{amssymb,stmaryrd}

\newcommand{\Z}{\mathbb Z}

\newcommand{\C}{\mathbb C}
\newcommand{\R}{\mathbb R}

\newcommand{\Dirac}{ / \hspace{-0.5em}\partial}

\newcommand{\beq}{\begin{equation}}
\newcommand{\eeq}{\end{equation}}
\newcommand{\beqarr}{\begin{eqnarray}}
\newcommand{\eeqarr}{\end{eqnarray}}
\newcommand{\beqa}{\begin{eqnarray*}}
\newcommand{\eeqa}{\end{eqnarray*}}
\begin{document}
\thispagestyle{empty}

\title{Weyl geometric gravity and  ``breaking'' of electroweak symmetry}
\author{Erhard Scholz\footnote{Wuppertal University, Faculty of Mathematics and Natural Sciences, \quad  scholz@math.uni-wuppertal.de}}

\date{05. 04. 2011 }
\maketitle

\begin{abstract}
A Weyl geometric scale covariant approach to gravity due to  Omote,  Dirac, and Utiyama (1971ff)  is reconsidered. It can be extended to the electroweak sector of elementary particle fields, taking into account their basic  scaling freedom. Already Cheng (1988)    indicated that electroweak symmetry breaking, usually attributed to the Higgs field with a  boson expected at $0.1 - 0.3 \,TeV$, may be due to a coupling between Weyl geometric gravity and electroweak interactions.   Weyl geometry seems to be well suited for treating   questions of elementary particle physics, which relate to scale invariance and its  ``breaking''. This setting suggests the existence of  a scalar field boson at  the surprisingly  low energy of $\sim 1\, eV$. That may appear unlikely; but, as a payoff,   the acquirement of mass arises as  a result of  coupling  to gravity in agreement with the understanding of  mass as  the gravitational charge of fields.

\end{abstract}

\subsection*{1\quad  Introduction}
In the 1970s  M. Omote, R. Utiyama,  P.A.M. Dirac and others took up the idea of
 Jordan-Brans-Dicke theory to study gravitational Lagrangians  linear in scalar curvature $R$ but coupled to a scalar field $\phi$ and studied it in the framework of Weyl geometry \cite{Omote:1971,Omote:1974,Dirac:1973,Utiyama:1975I,Utiyama:1975II}.   During the following roughly two decades,  different views of how mass  might be ``generated'' due to a link between Weyl geometric gravity and symmetry breaking at the electroweak energy level were developed   \cite{Smolin:1979,HungCheng:1988,Drechsler/Tann,Drechsler:Higgs}. Cheng indicated that the scalar field of gravity could well play a Higgs-like role, once it was extended to  the weak isospin sector. He was not the only one to explore such a connection. In 
Brans-Dicke theory   such a link was looked for in cosmology, the interaction zone  of the ``colliding fields''  gravity and elementary particle physics  \cite{Kaiser:colliding,Kaiser:mass}.  Few authors  drew upon the resources of Weyl geometry and the accordingly modified theory of gravity. W. Drechsler and H. Tann did so, but  preferred to model the origin of mass by  introducing   a Lagrangian term which  breaks scale symmetry explicitly. 

For about ten years the topic seemed to be put aside, because of different theoretical interests in the mainstream of elementary particle physics and of gravitation  theory. Only recently the discussion   was reopened by proposals which foresee  a Higgs-like role for the weak isospin extended  scalar field  of Weyl geometry, by   introducing a second (real) scalar field $\sigma $ transforming under rescaling with the same weight $-1$ as $\phi$ \cite{Nishino/Rajpoot:2004,Nishino/Rajpoot:2007}. 
After the first year of taking data at the LHC, it may be the right time to reconsider the topic of a possible connection between electroweak symmetry ``breaking'' and gravity from a  Weyl geometric perspective.  That is the goal of this article.

In order to make it essentially self-contained, and because of different notational conventions in the literature, basic definitions and properties of the Weyl geometric generalization of Riemannian geometry are recalled. A short discussion follows of how Weyl scaling is to be understood as a localized version of classical scaling in the sense of \cite{Barenblatt:Scaling}  (section 2.1). 

A Weyl geometric version of  gravity,  slightly differing from  Brans-Dicke theory by its explicit usage of the Weylian scale connection, but otherwise in strong formal analogy to the latter, was  introduced and studied by M. Omote, R. Utiyama, P.A.M. Dirac and others in the 1970s. Its main structural ingredient is the Weylian scale connection $\varphi$ and its curvature $f=d\varphi$. Assuming a dynamical role for scale curvature $f$ leads to a mass term close to the Planck scale. That was observed  by R. Utiyama, taken up in 1988  by Cheng. It was also seen, apparently independently, by L. Smolin in 1979. In the result, the dynamical aspect of scale symmetry is ``broken'' shortly below the Planck energy scale, and the Weylian scale connection loses the character of a field in its own right considerably below it. 
 Geometrically, an integrable version of Weyl geometry remains at the laboratory level and for astronomically/astrophysically directly observable scales (section 2.2). 

The scalar field $\phi$ of scale weight $-1$ ensures that this does not  automatically imply a reduction of the geometric structure to Riemannian geometry, and of gravitation to Einstein gravity, although it is formally possible to choose the scale gauge such that Riemannian part of the metric alone survives and the scale connection vanishes ({\em Riemann gauge}). But scaling freedom of the theory  and, with it, scale covariance of quantities remain; scale invariant expressions can be derived from them.  Riemannian geometry  expresses   scale invariant quantities correctly only in the case of a trivial scalar field which is  constant  in Riemann gauge. A comparison with Einstein gravity is possible  in different gauges, in particular those corresponding to ``Jordan frame'' or  ``Einstein frame'' in  Brans-Dicke theory (sections 2.3, 2.4).

The next passage contains a short exposition of how Weyl geometric gravity can be extended to the electroweak sector of elementary particle fields, taking into account their basic  scaling freedom (section 3.1). The possibility, and naturality, of this extension is one of the reasons why one should not give up scaling freedom of geometry too  early and without good physical reason. It should not come as  too great a surprise, if the conceptual advantages of Weylian  geometry   over Riemannian one  (even in the integrable version of Weyl structures)  turns out to be of physical importance. Scale invariance and its  ``breaking'' (more pragmatically,  its reduction) is an essential ingredient of elementary particle physics.  Weyl geometry is ideally suited to the purpose of giving such questions a clear conceptual and mathematical foundation, at least on  the classical level. In this respect it agrees with the intention of some recent studies in conformal field theory \cite{Gover_ea:2009,Shaukat/Waldron:2010}.

 Sections 3.2 and 3.3 discuss what happens if the ordinary Higgs mechanism (in its  
non-quantized form) is transferred from special relativistic fields to  Weyl geometrically ``curved'' spaces.  That leads naturally to the question whether, from the present point of view,  ``gravity can do'' what is usually  ascribed to   the Higgs ``mechanism'' (section 3.4). The  answer  indicated by Cheng, later roughly reiterated  by van der Bij \cite{vanderBij:1994},  is that theoretically it can.\footnote{Later van der Bij changed his mind.}
This article investigates how that can be done and what it  would mean. A most surprising aspect  of the answer is  that the scalar field boson, if any, would have classically negligible mass  far below the threshhold of present collider experiments, even after quantum corrections.  
 
 The core of this  analysis is  the biquadratic potential $R |\phi|^2 + \lambda _4 |\phi|^4 $ of the Weyl geometric scalar field, given by its coupling to scalar curvature $R$ (negative in the signature chosen  here) and the quartic scale invariant term with coefficient $\lambda _4$ substituting the old ``cosmological constant'' term of Einstein gravity.    The consequences of the potential for long range gravity are quantitatively tiny on solar system level, but of structural importance for cosmology. A most crucial difference to present standard cosmology is that, in the Weyl geometric framework, scalar curvature induced from the warp function of Friedman-Lemaitre cosmology becomes part of the field theoretic structure of the gravitational vacuum. Thus not only $\lambda _4$ but also $R$ induced from seemingly cosmological properties of large scale solutions should be present in empty space regions, otherwise well approximated by the classical Schwarzschild solution, and be attributed to  the gravitational vacuum (section 4.2). 
 
 Short commentaries of different views of electroweak ``symmetry breaking'' and on the still wide open question of the relationship between gravity and the quantum  round off the last section of the article (sections 4.1, 4.3).

\subsection*{2 \quad Weyl geometric gravity (Omote-Dirac approach)}
\subsubsection*{2.1 \quad Weyl geometry/Weyl structures}
Weyl geometry presupposes  a (differentiable) manifold with a  Weylian metric arising from a generalization of the Riemannian case. This generalization may be  motivated by a conceptual analysis of what a differential geometrical structure should be able to express in agreement with field theoretical principles, and what it should not  (no direct comparison of distant measurements).  From a purely structural point of view, the definition of a Weylian metric may be streamlined by the more recent concept of a Weyl structure which, however,  is not further considered here.\\[0.3em]
{\bf Weylian  metric, Weyl scaling.} \quad 
Weyl  introduced his  ``purely infinitesimal'' generalization of Riemannian geometry  as an   attempt to find deeper roots for Einstein's theory of gravity and to unify it with electromagnetism  \cite{Weyl:InfGeo,Weyl:GuE,Weyl:GuE_English}. Weyl's geometrical innovation was taken up in physics \cite{Pauli:1921}, not only  in  gravity \cite{Eddington:Relativity,Bergmann:Relativity,Dirac:1973,Israelit:Book,Adler/Bazin/Schiffer,%
Blagojevic:Gravitation}, but also in classical (unified) field theory \cite{Vizgin:UFT}. More recently, a handful of authors have considered  it as a  geometrical framework for exploring  connections between   electroweak theory and gravity \cite{Smolin:1979,HungCheng:1988,Drechsler:Higgs,Drechsler/Tann,Nishino/Rajpoot:2004}. In mathematics, Weyl geometry fared   less favourably. In the first decades after its invention few differential geometers took it seriously. Only with the rise of modern differential geometric methods after the 1950s, it started to attract geometers' interest \cite{Folland:1970}. Recently, Weylian scale gauge has been taken up and developed   not only for the Riemannian case, but also for complex (hermitian and K\"ahler) and quaternionic geometry, under the heading of {\em Weyl structures} \cite{Higa:1993,Gauduchon:1995,Calderbank:2000,Ornea:2001}.\footnote{Needless to say that the literature citations are far from complete. They are intended to give an exemplary orientation on the topics mentioned.}
As background literature physicists may like to use \cite[chap. 15.2]{Adler/Bazin/Schiffer} and \cite[chap. 4]{Blagojevic:Gravitation} and the classics
 \cite{Weyl:GuE_English,Eddington:Relativity,Bergmann:Relativity}; more mathematically inclined readers may prefer \cite{Weyl:InfGeo,Folland:1970,Higa:1993}. As conventions and notations differ in the literature,  we give a short account of basic concepts of Weyl geometry as  used here in a fairly informal way.
 
Weyl geometry works in a {\em Weylian manifold} $(M, [g, \varphi])$, or above it, as far as fibre bundle constructions are concerned.   $M$ denotes a  differentiable manifold, here usually of dimension $n:= dim \, M = 4$. $[g,\varphi]$,  the {\em Weylian metric},  is an  an equivalence class of pairs,  each  one consisting  of a (pseudo-) Riemannian metric $g$,   the {\em Riemannian component} of the metric, and a real valued differentiable 1-form $\varphi$, the   {\em scale} or {\em length connection}. Locally $g$  and $\varphi$ are represented by $(g_{\mu \nu}) = g_{ \mu \nu }dx^{\mu}dx^{\nu} $,  respectively  $ (\varphi_{\mu })=  \varphi_{\mu } dx^{\mu}$.  
 An equivalence between pairs $(g, \varphi) \sim (\tilde{g}, \tilde{\varphi})$ is given by    { conformally rescaling}  the Riemannian component of the metric
\beq \tilde{g}= \Omega^2 g  \, , \label{rescaling} \eeq
$\Omega >0$  a nowhere vanishing real valued function on $M$. Simultaneously the length connection has to be  {\em gauge transformed} as part of the same procedure 
\beq \tilde{\varphi} = \varphi - d \log{\Omega }.   \label{scale transformation} \eeq
 Choosing a representative  $(g,\varphi)$ of the equivalence class $[g,\varphi]$,  globally or locally, means to ``gauge'' the metric mathematically. Physically,  such a choice  expresses  a locally dependent introduction of length units $e_L$. 

We thus adopt a primarily {\em passive}  interpretation of Weyl scaling (transformation of units), because material structures usually do not allow  active up- or downscaling, at least not without further additional considerations \cite{Barenblatt:Scaling}. Historically, the revival of Weyl geometry in physics of the late 1960s and early 1970s was triggered by   experimental indications of an  approximative active scaling symmetry of deep inelastic scattering, i.e., form factors remained nearly  invariant  under rising energies. Further refinement of experimental knowledge has shown that, also in high energy physics, active scaling symmetry is broken and only of   rough approximative validity. Nevertheless, in order to understand what is ``broken'' and how, respectively why, we need a deeper understanding of the underlying structure linking geometry and field theory. For that a mathematical framework  designed to incorporate scale symmetry is useful. In this sense the reasons leading to the revival of Weyl geometry in physics continue to be valid, although in a different overarching  perspective, more modest than in the 1960s. A similar motivation seems to lie at the basis of a recent  research program studying the relation between Weyl's gauge invariance, conformal geometry and mass generation using the symbolic tools of  ``tractors'', an extension of the tensor concept adapted to conformal rescaling \cite{Gover_ea:2009,Shaukat:Diss,Shaukat/Waldron:2010}.  This approach differs from the Weyl geometric one by using a  peculiar kind of covariant gradient operator (``Thomas $D$-operator'') which is no covariant derivative but can be ``employed to a similar effect'' \cite[428]{Gover_ea:2009}.\footnote{The Thomas $D$-operator is defined by means of the Levi-Civita derivative of any of the scaled metrics of the conformal class. It operates on  quantities which transform according unified rules under scale transformations (tractors). It serves the purpose of unifying calculations in a conformal structure. In this way it differs from Weyl geometry proper.}

If we add  a { global convention} for the value of the velocity of light $c$ and the Planck constant $\hbar$ to the local choice of the length unit $e_L$,  physical units are determined according to the principles of dimensional analysis and classical scaling \cite{Barenblatt:Scaling}. 
 In particular, the units for time $e_T$ and energy $e_E$ become $e_T:= c^{-1}\, e_L$,  $e_E:= \hbar \, e_T^{-1}$ etc.\footnote{Of course, in place of a the local choice of length units also energy units might be  chosen locally as the basic physical quantity. From a phenomenological point of view, the scale connection $\varphi$  then represents an ``energy connection'' rather than a ``length connection'' like in Weyl's original view. Mathematically both views are  equivalent.}
 The total procedure, based on local scaling of length (or energy) units and a global convention  for the numerical values of $c$ and $\hbar$, will be called   {\em Weyl scaling} of physical units. 

As $c$ and $\hbar$ are  dimensionally independent physical constants, they can be given arbitrary  numerical values $|c|, |\hbar|$ also in the Weyl geometric context, like in ``classical'', i.e. global, dimensional analysis \cite{Barenblatt:Scaling}. Typical examples for such global conventions are, e.g., 
\begin{itemize}
\item[(1)] $|c|=|\hbar|:=1$ (``natural units''), 
\item[(2)] or $|c|:= 2.99792458\cdot 10^{10}, |\hbar|:= 6.5821220\cdot 10^{-16}$ (comparable to the values in the cgs system).
\end{itemize}
In both examples no numerical value of the gravitational constant has been prescribed; attempting to do so would break Weyl scaling freedom/symmetry. Therefore the ``natural'' units (1) are not identical to Planck units and the  system (2)  is not identical with  cgs units. An appropriate specification of legth/time or energy units in addition to (2) leads to the latter (cgs).\footnote{In or context, a natural choice  in scalar field gauge is  (\ref{v}).} 
  
 In the physics literature, the scale connection $\varphi =(\varphi_{\mu})$ is often called ``the Weyl (co-)vector field''. This  is a misleading terminology;  one has to keep in mind that $\varphi$ transforms  as a  connection with values in the (multiplicatively trivial) Liealgebra $\R$ of the localized group $(\R^+, \cdot)$, rather than as a covector field in the proper sense. 
 
In general, we assume  Lorentz signature of the Riemannian part of the metric, $sig\, g = (1,3)=(+---)$. Geometrical investigations in general relativity often prefer signature $(3,1)=(-+++)$. For translation between the signature conventions we use the factor  $\epsilon_{sig}=+1$ for $sig\,( g) = (1,3)$, and $\epsilon_{sig}=-1$ otherwise.\\[0.5em]
 \noindent
{\bf Important properties (theorems)} \quad 
 A direct comparison of lengths of vectors $\xi , \eta$  is  meaningful only if  they  are ``attached'' to the  same point of the manifold, $\xi , \eta \in T_p M $; similarly for other physical quantities  which are affected by local choice of scale, like energy, momentum etc. These will be called {\em scale covariant } quantities (affected by the choice of the Riemannian component of the Weylian metric) or  {\em Weyl fields}.   If a  (scalar, vector, tensor spinor) field $X$ on $M$ is   affected by rescaling  under $\Omega $ according to (\ref{rescaling}) such that
$   X   \mapsto \tilde{X}= \Omega ^k X$ with $k \in \R \; ,  $ 
then  $w(X):= k$ is  the Weyl or scale  {\em  weight} of $X$.
  
 Comparison of scale covariant quantities $X(p),  X(q)$ at different points $p \neq q$ is meaningful only  after ``transport of length standards'', resulting in  multiplication of $X(p)$ by  $\lambda (\gamma )^{w(X)}$, where $\lambda(\gamma ) $ is a ``scale transfer'' function arising from integration of the scale connection along a path $\gamma $ between $p$ and $q$ 
\[  \lambda (\gamma ) = e^{\int_{\gamma } \varphi (\gamma ') } .\] 
Important theorems of Weyl geometry are:
\begin{itemize}
\item Like in the Riemannian case, a Weylian manifold $(M,[g,\varphi])$ possesses a uniquely determined  compatible affine (i.e., torsion free)  connection $\Gamma = (\Gamma ^{\mu }_{\nu \lambda } )$,  the  Weyl geometric {\em Levi-Civita connection} with corresponding {\em covariant derivative} $ \nabla = \nabla_{\Gamma }$, compatible with the Weylian metric.\footnote{Compatibility with the metric may be expressed by the demand that  parallel transport of a vector $X(p)$ along a path $\gamma $ from $p$ to $q$ to $X(q)$ leads to consistency with length transfer: $|X(q)|= \lambda (\gamma )  | X(p)| $. \label{fn compatibility}} 
\item ``Uniquely determined'' means  that $\Gamma$ and  $ \nabla = \nabla_{\Gamma }$ are  independent of scale gauge; in particular,  $\nabla$ is a ``scale invariant'' covariant derivative.

\item Thus also the Riemannian {curvature} $Riem = (R^{\mu}_{\lambda \nu \kappa })$ of $\Gamma$ and its Ricci curvature $Ric=( R_{\mu \nu })$ are  scale invariant, i.e., independent of the choice of units. Scalar curvature $R = R^{\lambda }_{\lambda }$, on the other hand,  is  only{ scale covariant} and of weight $w(R)=-2$.
\item The  Weyl geometric Levi-Civita connection $\Gamma $ can be expressed in terms of the Levi-Civita connection $_g\hspace{-0.0em}\Gamma $ of the Riemannian component of any gauge $(g,\varphi)$ and of the corresponding scale connection $\varphi$ by  
\beq  \label{Christoffel}  \Gamma^{\mu }_{\nu \lambda } =   {}_g\Gamma^\mu _{\nu \lambda } + \delta ^{\mu }_{\nu } \varphi _{\lambda } +
\delta ^{\mu }_{\lambda } \varphi _{\nu } - g_{\nu \lambda } \varphi^{\mu } \, . \eeq 
\item Similar reductions exist for the Weyl geometric  curvatures  $Riem, Ric, R$.\footnote{Here the sign convention of differential geometry is used, $Riem (X,Y)Z= \nabla_X\nabla_Y Z- \nabla_Y \nabla_Y Z - \nabla[X,Y] Z$; in Ricci calculus $R^{\nu }_{\mu \lambda \kappa} = \partial _{\lambda }  \Gamma ^{\nu }_{\mu \kappa } -  \partial _{\kappa }  \Gamma ^{\nu }_{\mu \lambda   }+ (\Gamma^{\rho }_{\mu \kappa } \Gamma^{\nu }_{\rho \lambda } - \Gamma^{\rho }_{\mu \lambda } \Gamma^{\nu }_{\rho \kappa })$. Contraction convention for $Ric$ is $Ric_{\mu \nu }=R_{\mu \nu }=R^{\lambda }_{\mu \lambda \nu }$.}
 They can be expressed, in any gauge, by corresponding curvatures of the Riemannian component $_{g}\hspace{-0.1em}R $  and expressions in the scale connection $_{\varphi}\hspace{-0.1em}R $ etc. In particular 
 $Ric = _g\hspace{-0.3em}Ric + _{\varphi}\hspace{-0.3em}Ric $, $ R = _g\hspace{-0.3em}R+ _{\varphi}\hspace{-0.3em}R $ with
 \beqarr
 _{\varphi}\hspace{-0.1em}Ric_{\mu \nu} &=&  (n-2)(\varphi_{\mu}\varphi_{\nu}   - _g \hspace{-0.2em}\nabla_{(\mu } \varphi_{\nu)}) - g_{\mu \nu}((n-2)\varphi_{\lambda } \varphi^{\lambda }   +   _g \hspace{-0.25em}\nabla _{\lambda }\varphi^{\lambda }) \nonumber \\
  (\mbox{for}\; n=4) &=&  2 (\varphi_{\mu}\varphi_{\nu}   - _g \hspace{-0.2em}\nabla_{(\mu } \varphi_{\nu)}) - g_{\mu \nu}( 2 \varphi_{\lambda } \varphi^{\lambda }   +   _g \hspace{-0.25em}\nabla _{\lambda }\varphi^{\lambda }) \;  ,
\label{Ric} \\
_{\varphi}\hspace{-0.1em}R &=& - (n-1)(n-2) \varphi_{\lambda } \varphi^{\lambda }  - 2(n-1) _g \hspace{-0.2em}\nabla _{\lambda }\varphi^{\lambda }   \nonumber \\
(\mbox{for}\; n=4) &=& - 6 \varphi_{\lambda } \varphi^{\lambda }  - 6\,  _g \hspace{-0.2em}\nabla _{\lambda }\varphi^{\lambda } . \label{R}
 \eeqarr
Here $_g \hspace{-0.2em}\nabla$ denotes the covariant derivative with respect to the Riemannian component only,   $_g \hspace{-0.2em}\nabla = \nabla_{ _{g \hspace{0em}}\Gamma}$.

\item   {\em Scale curvature}, i.e. the curvature  $f$ of the scale connection,  is $f = d \varphi$. Note language: scale curvature $\neq$ scalar curvature ($=$ ``Ricci scalar''). 
\item If $f=0$, there exist a local choice of units s.th. $ \tilde{\varphi}=0$. 
In this case we have, in simply connected regions,  a locally  {\em  integrable  Weyl  geometry}  (IWG).
Then the scale  choice  $(\tilde{g},0)$  defines a Riemannian metric. It  is called  {\em Riemann gauge} of the IWG. 

\item Generally, the covariant derivative $\nabla X$ of a scale covariant quantity $X$ is, itself, not scale covariant. Therefore  a {\em scale covariant derivative} (sometimes also called ``Weyl covariant derivative'') $D$ is defined by
\beq DX := \nabla X + w(X)\varphi  \otimes X \eeq
 $DX$ is again scale covariant (a Weyl field) and $w(DX)=w(X)$.\footnote{Compatibility of $\nabla$, respectively $D$, can now be expressed l by the condition that in every gauge $D_X g(Y,Z)=g(D_X Y,Z) + g(Y,D_X Z)  $  for any covariant vector fields $X,Y,Z$.} 
\item Example: $\nabla g$ is not scale covariant, but $Dg$ is. 
Moreover 
\beq Dg=\nabla g + 2  \varphi \otimes g=0 \; ; \label{semimetricity} \eeq  i.e., $g$ is { scale covariantly constant}. Compare: In Riemannian geometry the metric is covariantly constant, $_g{\hspace{-0.3em}\nabla}  = 0$. 
\item In the literature of theoretical physics (\ref{semimetricity}) is sometimes considered as an indicator of ``non-metricity'' of the scale covariant derivative; sometimes it is called, more generously,  ``semi-metricity''. From a  Weyl geometric point of view, the scale covariant derivative is just ``metrical'', $Dg=0$. 
\end{itemize}
{\bf Comparison of conventions}\quad 
In the physics literature (and in Weyl's writings) the weights are often half of ours, because the scaling condition is written as  $\tilde{g} = \Omega\, g$ rather than $\tilde{g} = \Omega^2 \, g$ . Moreover, in most of the physics literature (not so in most of Weyl's writings) the sign convention for the scale connection $\varphi_{lit}$ is inverse to ours (and Weyl's generic choice); both together result in $\varphi_{lit}= - 2 \varphi $. Thus the formulas (\ref{scale transformation}) ff. appear in a slightly different form,  (\ref{scale transformation}) itself becomes, e.g.,  $\tilde{\varphi_{lit}}= \varphi_{lit} + \frac{1}{2} d \log {\Omega_{lit}}$ etc. ($\varphi$ our convention,  $\varphi_{lit}$ the other one). The reason for our choice of  sign can be found at the end of  footnote \ref{fn compatibility} (in addition to keeping to Weyl's generic convention); the weights are oriented at lengths as  reference quantities.\footnote{It even may happen that  both weight conventions are used by the same author. In the otherwise  rich and scholarly work \cite{Blagojevic:Gravitation}, both conventions even appear  in one and the same chapter, ours in  4.2 (``Weyl-Cartan geometry''), the other one in 4.1 (``Weyl gauge invariance'').}\\[0.5em]

\subsubsection*{2.2 \quad Weyl geometric gravity,  Omote-Utiyama-Dirac approach}
 Weyl geometry is a modest modification of Riemannian geometry; therefore it allows for  natural   generalizations  of Einstein gravity. Its basic principle is scale invariance of the underlying Lagrange density $\mathcal{L} = L \sqrt{|det\, g|}$. As $\sqrt{|det\, g|}$ is of weight 4, only Lagrange functions $L$ of weight $w(L)=-4$ are to be considered, in particular for the gravitational term $L_{G}$. Weyl explored Lagrangians $L_{G}$ quadratic in the curvature.  Coupling of curvature with a scalar field  broadens the possibilites. The first author to apply such a Brans-Dicke like modified Hilbert action in Weyl geometry was M. Omote \cite{Omote:1971}. A little  later, papers  by Utiyama and Dirac followed \cite{Dirac:1973,Utiyama:1975I,Utiyama:1975II}.\footnote{Utiyama knew Omote's paper and cited it, while Dirac was silent about it. I owe  the hint at Omote's work to F. Hehl. }
 For the sake of conciseness, I shall call that approach {\em Weyl-Omote-Dirac gravity} (WOD).
For an excellent survey see \cite[chap. 4]{Blagojevic:Gravitation}.\\[0.5em]
 \noindent
 {\bf Lagrangians of Weyl-Omote-Dirac gravity (WOD)} \quad
 The  action of a Weyl geometric real or complex scalar field $\phi$ of weight $-1$ contains the kinetic  Lagrangian density $\mathcal{L}_{\phi}=\gamma L_{\phi}\sqrt{ | det \,  g|}$  with constant  $\gamma $ and
 \beq L_{\phi} = \frac{1}{2} D_{\nu } \phi^{\ast} D^{\nu } \phi  \, , \label{L_phi}
 \eeq
  where $D$ is the scale covariant derivative. For  a scalar field,  $D_{\nu } \phi = \partial_{\nu}\phi - \varphi_{\nu} \phi$.  $\phi^{\ast}$ denotes complex conjugation (later, in extensions, it will denote dualization in complex vector spaces or an even more general conjugation).  
 
The  gravitational Lagrangian density $\mathcal{L}_{G\phi}  = L_{G \phi} \sqrt{ | det \,  g|}$ couples scalar curvature in a Brans-Dicke like manner to  the Weyl geometric   scalar field $\phi$. Bringing its own coupling factor $\gamma $  in order of magnitude into agreement with the one of the scalar curvature (by putting $\gamma = \xi ^2$), we can write 
 for $sig \, (g) = (1,3)$:\footnote{Change of signature flips all signs (take care for $V_4$ !), variation leads to the same dynamical equations.}
\beq  L_{G \phi} = \frac{\xi^2}{[\hbar c]} (- \alpha  |\phi|^2   R  + \frac{1}{2} D_{\nu } \phi^{\ast} D^{\nu } \phi -  V_4 (\phi)) \label{L_G_phi}
\eeq
 $\alpha $ is a coupling constant, regulating relative strength of scalar field kinetic term and the modified Hilbert term;  $\xi^2$ (and with it $\gamma $) is a  squared ``hierarchy factor'' (see section 3, equ. (\ref{hierarchy factor})), $|\phi|^2 = \phi^{\ast}\phi$.
 For $\alpha = -\frac{1}{12} $ one gets Penrose's conformal coupling in $dim \, M=4$.  Many authors choose this value, among them Dirac \cite{Dirac:1973} and his followers,  e.g.
 \cite{Israelit:Book,Drechsler/Tann,Drechsler:Higgs}. 
In our context, that is  an  unnecessary restriction, as far as $L_{G\phi}$ is concerned.\footnote{Considering the (abelian) Yang-Mills action $-\frac{1}{4}f_{\mu \nu }f^{\mu \nu }$  of $\varphi$, a reasonable although not compelling motif for this choice becomes visible; see below.}
 Weyl geometric scalar curvature $R$ and the scale covariant derivatives $D$ assure weights $-4$ for all terms of $L_{G\phi}$ anyhow, and thus scale invariance of  $\mathcal{L}_{G\phi}$.  We may just as well leave $\alpha $ open, at least for the moment. In doing so, we follow \cite{Omote:1971,Omote:1974},    presented also in \cite[102]{Blagojevic:Gravitation},  and (with the restriction $\alpha = 1$) in  \cite{Utiyama:1975I,Utiyama:1975I,Smolin:1979}.\footnote{Because of his interest in quantization, Smolin used a systematially broader range of all the possible scale invariant curvature terms, including those of  second order. Utiyama was aware of Omote's work and quoted it. Smolin did not; neither did he quote  Dirac.} 
 
$V_4$ is a quartic polynomial written in a form such that it changes sign with signature change
\[ V_{4} (\phi)  =  \epsilon _{sig}\,  \sum_{j=1}^4  \lambda_j |\phi|^j     \, . \] 
 Scale invariance demands $\lambda_j=0$ for $j=1,2,3$; but some authors decide in favor of an explicitly symmetry breaking term $\lambda _2 \neq 0$ \cite{Drechsler/Tann}\footnote{A similar explicit breaking by a quadratic term was  already foreseen in a setting of  conformal coupling between scalar curvature and the scalar field by \cite{Deser:1970}.} or introduce a second (real) scalar field $\sigma $ of weight$-1$, setting $\lambda _2=\sigma ^2$ \cite{Nishino/Rajpoot:2004}. We consider this  artificial and try to avoid it, keeping  the simplest possible scale invariant  Lagrangian, as long as  we are not  compelled  to give it up by problems of empirical adequacy. That reduces the potential to a  quartic monomial, 
 \beq  V_{4 } (\phi)  =  \epsilon _{sig}\,     \lambda_4 |\phi|^4 \; .  \label{L_V4} \eeq 
$V_4$ plays the formal role of a scale invariant ``cosmological'' term. Its energy tensor will  be absorbed by the energy tensor of  the scalar field more  broadly (see below, equ. (\ref{Theta})).
 
Assuming a dynamical role for the Weylian scale connection,   a Maxwell-like Yang-Mills action
$ \mathcal{L}_{\varphi} = L_{\varphi}  \sqrt{ | det \,  g|} $ has to be added with
\beq L_{\varphi} = - \frac{[\hbar c]}{4 }f_{\mu \nu}f^{\mu \nu} \, .   \label{L_varphi}\eeq

Disassembling the modified Hilbert term according to (\ref{R}) and  the kinetic term of $\phi$, leads to a Proca mass-like term for the scale connection
$ \frac{ \xi^2 }{[\hbar c]} (6\alpha +\frac{1}{2}) \varphi_{\mu} \varphi^{\mu}$, the  coefficient of which should be $ = \frac{1}{16 \pi G} $ in the GRT ``limit''. The field is massless iff $\alpha =-\frac{1}{12} $. In all other cases the scale connection acquires a huge mass.
The Proca mass factor turns out to be $\frac{1}{2} \frac{(m_{\varphi}c^2)^2}{\hbar c} =\frac{c^4}{16 \pi G} $ and thus close to the Planck scale  \cite{Smolin:1979,HungCheng:1988}:
 \beq m_{\varphi}^2 = \frac{\hbar c}{8 \pi G}  = \frac{1}{8 \pi} m_{Pl}^2 \;  \label{Proca mass}\eeq
 This surprising effect has been seen and discussed by Utiyama, Smolin and later  again  by Cheng.\footnote{A misplaced sign choice  led Utiyama even to assume a ``tachyonic'' Planck mass. That was corrected by \cite{Hayashi/Kugo:Weyl_field}.}  Probably Dirac chose the value of $\alpha =-\frac{1}{12} $ deliberately (using our notational conventions) in order to avoid such mass effects of the scale 
connection.\footnote{He did not mention that, but rather  talked about getting ``the simplest equation for the vacuum''. Moreover, he sticked to the interpretation of $\varphi$ as electromagnetic potential, long before given up by Weyl \cite[410]{Dirac:1973}.}
Here we are not interested in the details of this discussion. The common structural features in the papers of the authors mentioned   are characterized by the combined Lagrangians (\ref{L_G_phi}) + (\ref{L_varphi}).

The second long range interaction field,  classical electromagnetism with 4-potential $i A=i (A_{\mu})$, has the well known scale invariant  (in flat space even conformally invariant) Lagrangian 
\beq  L_{em}= - \frac{[\hbar c]}{4} F_{\mu \nu} F^{\mu \nu} \, ,\eeq
with $F^{(em)}=(F_{\mu \nu}) = d A$.

Lagrangians of classical matter fields $\mathcal{L}_m$ may  be adapted to a scale covariant form, by modifying their action densities  as demanded by dimensional analyis and principles of Weyl scaling of units, e.g. like in   \cite{Scholz:FoP}.  An extension to fermionic fields $\mathcal{L}_{\Psi}$ and the  electroweak sector of bosonic interactions $\mathcal{L}_{ew}$ of the standard model of particle physics will be discussed in  section 3. \\[0.5em]
\noindent
{\bf Arguments for the  integrable version, IWOD} \quad 
As the  range of the  field strength of the scale connection is restricted  to the order of the Planck length  $l_{Pl}$ (if it represents a dynamical entity at all), $\varphi$  will not be able to display any observable curvature effects far above  $l_{Pl}$. For all ``practical purposes'', far below the Planck energy, respectively far above the Planck length,  we can safely consider $f$  and  the Lagrangian (\ref{L_phi}) as negligible. With the  Lagrangian (\ref{L_G_phi})  alone we effectively  work in  {\em integrable Weyl-Omote-Dirac (IWOD) geometry}. Only coming close to the Planck scale, one has to consider the full Lagrangian (\ref{L_G_phi})+(\ref{L_varphi}) and ``full'' Weyl geometry  with non-vanishing scale curvature $f$.  

These structural properties seem to indicate 
 that the scale symmetry of integrable WOD gravity may display  a non-dynamical residuum of a truely dynamical symmetry which becomes effective  close to the Planck scale. If this is the case, it may turn out that $\phi$ can be considered  an order parameter of some condensate after ``spontaneous breaking of dynamical scale symmetry'' close to the Planck scale.  Given our present ignorance of Planck scale physics, it would be premature, however, to delve deeper into such speculative waters here. For our purpose, we can perfectly well work in the mathematical structure of integrable Weyl geometry, independent of  further ontological interpretations. 
\\[0.5em]
\noindent
{\bf Dynamical equations, in particular Einstein  equation} \quad
Assuming a total Weyl geometric Lagrangian of the form
\[  \mathcal{L}_{G\phi} + \mathcal{L}_{M}+ \mathcal{L}_{Int} \]
with classical or field theoretic matter terms indexed by $M$ ($ m$, respectively $\Psi$) and interaction terms indexed by $Int$ (electromagnetic oder standard model interactions),  dynamical equations are derived as usual  by variation $ \delta _{{v}}$ with respect to the dynamical variables ${v}$. In particular, variation with respect to the Riemannian component of the metric $g$ leads to a Weyl geometric  version of the Einstein equation.\footnote{See, among others, \cite{Tann:Diss,Drechsler/Tann,Blagojevic:Gravitation}}

Varying the modified Hilbert term in $\mathcal{L}_{G\phi}$ with respect to $g$ leads to  an expression similar to the classical one plus an additional term $A$,
\beq \sqrt{|det\,g|}^{-1} \delta _{g}( |\phi|^2 R) = |\phi^2| (Ric - \frac{R}{2}g ) + A  \eeq
with $A_{\mu \nu}= D_{\lambda} D^{\lambda} |\phi |^2 g_{\mu \nu} - D_{(\mu}D_{\nu)}|\phi|^2  $ \cite[(372)]{Tann:Diss}, \cite[(4.47)]{Blagojevic:Gravitation}. $A$ has the form of an additional contribution to  the dynamical energy tensor of the scalar field $T^{(\phi)}$ and is often called an ``improvement'' term of it.\footnote{It has been noted by different researchers that $A$ contributes to the energy tensor of the scalar field in exactly the same way as the ``improvement'' term of  Callan, Coleman and Jackiw \cite{Tann:Diss,Drechsler/Tann}, \cite[96ff.]{Blagojevic:Gravitation}.}
The dynamical energy momentum tensor  is generated by taking the variational derivative  of   the Lagrangian $\mathcal{L}_{\phi}$, like for any matter or interaction field $X$, 
\beq
T^{(X)}= 2 \sqrt{|det\,g|}^{-1} \delta _{g} \mathcal{L}_{X} \, , \label{T^(X)}
\eeq
\[ \sqrt{|det\,g|}^{-1} \delta _{g} \mathcal{L}_{\phi}   =  \frac{\xi^2}{2}  (D_{(\mu}\phi^{\ast}D_{\nu)}\phi    -\frac{1}{2} D_{\lambda}\phi ^{\ast}D^{\lambda} \phi   \, g_{\mu \nu} )\, . \]

 With  $A$ and the contribution from the  $V_4$ term, $|det\,g|^{-1} \delta _{g}(-V_4)=\frac{V_4}{2}g $, the 
 {\em Weyl geometric Einstein equation} in the {\em massless} case $ \mathcal{L}_{M}+ \mathcal{L}_{Int} =0 $, becomes:
 \beqarr   Ric - \frac{R}{2}g &=&  \Theta^{(\phi)} =  \Theta^{(I)}+  \Theta^{(II)} \label{massless Einstein equation} \\
     \Theta^{(I)} &=&   |\phi|^{-2}( \frac{1}{2 \alpha }V_{4 }   - \frac{1}{4\alpha } D_{\lambda}\phi ^{\ast}D^{\lambda} \phi  -  D_{\lambda} D^{\lambda} |\phi |^2 ) \, g \label{Theta} \\
        \Theta^{(II)}_{\mu \nu} &=&  |\phi|^{-2}\left(  \frac{1}{2 \alpha } D_{(\mu}\phi^{\ast}D_{\nu)}\phi +   D_{(\mu}D_{\nu)}|\phi|^2 \right) \nonumber 
 \eeqarr 
with Weyl geometric curvatures $Ric, R$ and scale covariant derivatives $D_{\mu}$.
 Terms without factor  $\alpha ^{-1}$ are due to the ``improvement'' $A$.
 
$  \Theta^{(I)}$ is proportional to the Riemannian component of the metric, i.e., of the form  $\tilde{\Lambda}g$ with $\tilde{\Lambda}=  |\phi|^{-2}( \frac{1}{2 \alpha }V_{4 }   - \frac{1}{4\alpha } D_{\lambda}\phi ^{\ast}D^{\lambda} \phi  -  D_{\lambda} D^{\lambda} |\phi |^2 )$ (in any scale gauge). In this sense, it looks like the vacuum or ``dark energy'' tensor $\Lambda g$ of the received approach. Note, however, that in our case the negative pressure $-\tilde{\Lambda}$ is {\em no constant}, independent of the matter content of the universe, but dynamically determined by the scalar field. The latter is, in turn, related via curvature to  matter  by the scalar field equation (below) and the potential condition (section 3, equ. (\ref{potential condition}).  

$\Theta^{(II)}$ is an additional energy-stress-momentum contribution of $\phi$, beyond the metric proportional $  \Theta^{(I)}$.  It may be worthwhile to investigate whether there are  cosmologically relevant solutions in which the matter content is  dominated by  $\Theta^{(II)}_{00}$; if so, that might open   new vistas on  the ``dark matter'' problem.

If massive fields or non-gravitational interactions are present,  $ \mathcal{L}_{M}+ \mathcal{L}_{Int} \neq 0 $, the energy momentum tensors $T^{(X)}$,  $ X=M, Int$, appear as additional terms on the right hand side of the {\em Einstein equation} with {\em massive} source:
\beq Ric - \frac{R}{2}g =  \Theta^{(\phi)} + \frac{[\hbar c]}{2 \alpha \xi ^2 |\phi|^2} (T^{(M)}+T^{(Int)}) \label{massive Einstein equation}
\eeq

Variation with respect to $\phi^{\ast}$ leads to  the  {\em scalar field equation}\footnote{Compare, among others, \cite{Drechsler/Tann,Blagojevic:Gravitation}.}
\beq  D_{\nu}D^{\nu}\phi = - \left(2 \alpha R+ 4\epsilon _{sig} \lambda _4 |\phi|^2   \right) \phi \, + \mbox{terms in $\delta _{\phi^{\ast}} {L}_M $}.
\label{scalar field equation}
\eeq
If  the Higgs-like potential condition for the ground state of the scalar field,  discussed in the next section is realistic, the scalar field equation governs only the classical field fluctuations about the ground state.  

For $\alpha \neq -\frac{1}{12} $ the expectation of a  massive boson near the Planck scale reduces the range of the field strength $f$,  and  the effective curvature effects of the scale connection $\varphi$, to Planck scale physics. At larger distances $\varphi$ is effectively integrable and thus reduced to its   purely geometrical role as part  of the metrical structure of IWG.\footnote{This difference is  analogous to Riemann's observation, in the third part of his inaugural lecture, that space curvature may effectively vanish on large scales, while there may be strong curvature effects in the small. He warned that one ought to be 
 cautious in drawing conclusions  from astronomical evidence for vanishing sectional curvature: ``\ldots we cannot draw conclusions from metric relations of the great, to those of the infinitely small; (\ldots) the curvature at each point may have an arbitrary value in three directions, provided that the total curvature of every measurable portion of space does not differ sensibly from zero''  \cite[68]{Riemann:HabilEnglish}. \label{fn Riemann} } 
 It  establishes consistency under scale transformations, comparable to the correction terms in $\partial \Omega $ of the affine connection in  Jordan-Brans-Dicke (JBD) theory.

 That is obviously different for the scalar field with its kinetic term (\ref{L_phi}) and its coupling to gravity. In the next subsection it will  become clear that in certain scale gauges, the dynamical properties of the scalar field $\phi$ may be expressed by the length connection $\varphi$. But even then the dynamical properties of both  are due to the scalar field rather than to the scale connection as such, which can be scaled away. 

The (second) Yang-Mills equation  for $\varphi$ with curvature $f=d \varphi=0$,  
\[ \ast\, d \hspace{-0.2em} \ast  \hspace{-0.2em}f = J \, , \]
becomes trivial sufficiently above the Planck scale. Varying with regard to $\varphi$  and  discussing  Yang-Mills theory in IWOD gravity would turn an idle wheel. It may become important for physics at the Planck scale only.  

\subsubsection*{2.3 \quad Scale invariant observables, scalar field gauge}

Freedom of scale choice seems to introduce an arbitrary element in the determination of metrical (scale covariant) quantities $X$. For a meaningful comparison of measured, or calculated, quantities, it is necessary to extract some {\em scale invariant magnitude} $\hat{X}$ of any  scale covariant $X$. In geometry this is an old problem, solved already in antiquity and later taken up by Descartes, through  the introduction of {\em proportion} theory and proportions of geometrical quantities.\footnote{\cite{Euclid:ElementsEnglish,Bos:Descartes}} 
In this way  ``real quantities'' of extended geometrical  magnitudes were made independent of the (subjective) choice of units. The same idea works, generalized, in Weyl geometry for any local physical quantity underlying Weyl scaling. One only needs to form proportions of the correct weight with some  scale covariant reference quantity. In WOD theory, the norm of the scalar field $|\phi|$  is a natural candidate for a universally accessible reference quantity.\footnote{ Weyl applied basically  the same idea with respect to scalar curvature as reference quantity for his gravitation theory  (second order in $R$)  \cite[298f.]{Weyl:RZM5}. } 
 Some authors therefore even declare $\phi$ to be a ``measure field'' \cite{Omote:1974,Utiyama:1975II,Israelit:Book}.

That leads to an obvious method for extracting a  {\em scale invariant magnitude} $\hat{X}$ from a scale covariant {\em local quantity} $X$ of weight $w:= w(X)$. One just has  to consider the  proportion with the appropriately weighted power of $|\phi|$ at any point $p$:
\beq
  \hat{X}(p) := \frac{X(p)}{|\phi|^{-w}(p)} = X(p)\, |\phi|^w(p) \;   \label{scale invariant magnitude}
\eeq
(the negative sign in the exponent of the denominator arises because we work with length, not energy, weights).
By definition $\hat{X}$ is {\em scale invariant}.

An experimentally {\em observable magnitude} $X^{(exp)}$ corresponding to a  magnitude $X$ of the theory is  measured by  a system of appropriate  material devices and scientific practices. The devices, e.g. radiotelescopes, or the  detectors of the LHC, are  lawfully constituted by classical systems (probably environmentally decohered from quantum states) and  are operated by a complex of technical practices, on which several levels of data selection and evaluation are superimposed, based on well defined theoretical practices. The corresponding  observable magnitude $\hat{X}(p)$   of the theory should basically be  proportional to 
 $X^{(exp)}$,     up to a global scaling factor depending on unit choice $const _u$:
\beq  \hat{X}(p)  =  const _u \, X^{(exp)}   \, \label{observable magnitude}  \eeq 

For any nowhere vanishing scale covariant quantity $Y$ of weight $w(Y)\neq 0$ there is a scale gauge in which $|Y|=const$; just set $\Omega = |Y|^{\frac{-1}{w(Y)} }$. In his early theory, Weyl proposed  to consider in a Weylian manifold with nowhere vanishing scalar curvature  the scale gauge in which  $R=const$ \cite[298f.]{Weyl:RZM5}. It will   be called {\em Weyl gauge}.\footnote{Not to be confused with ``Weyl gauge'' in electromagnetic theory.}

 In WOD gravity another specification appears much more natural,\footnote{Both scale gauges are closely related in WOD gravity; see section 3.3.}
  a scale gauge in which the scalar field is normed to a constant.  We call it {\em scalar field gauge}.  Starting from any gauge $(g,\varphi)$ in which the scalar field may assume the value $\phi$, the scalar field gauge $(\check{g}, \check{\varphi})$  is  given by
\beq \check{g} = \Omega ^2 g\, \qquad \mbox{with} \qquad \Omega := const\,  |\phi | \, . \label{scalar field gauge}
\eeq 
  In it,   the invariant magnitudes are directly expressed (up to a global constant factor)  by the scale covariant magnitude itself. 
 A scale covariant magnitude $X$ given in scalar field gauge will be denoted accordingly by $\check{X}$, a dotted equality $\doteq$  expresses relations which hold in  a specific gauge  only (preferably scalar field gauge).  Thus $\check{X} \doteq \beta \,  \hat{X}$, up to a constant factor $\beta $.

 \subsubsection*{2.4 \quad Two transitions to Einstein gravity} \qquad
 There are two natural transitions  to Einstein gravity. 
\begin{itemize}
\item  In IWOD gravity the scale connection can be integrated away. By geometrical default, one may thus like to choose Riemann gauge. Then the geometry looks like usual, but the graviational coefficent  $\frac{[\hbar c]}{2 \alpha \xi ^2 |\phi|^2}$ in (\ref{massive Einstein equation}) is no longer a constant but depends on the norm of the scalar field (comparable to ``Jordan frame'' in JBD theory). Moreover, the scale invariant magnitudes $\hat{X}$ can be calculated only if the norm of the scalar field is known!
\item If one takes the methodological principle of scale invariance seriously, the choice of scalar field gauge is  much more compelling (``Einstein frame'' in JBD theory). Then the scale invariant magnitudes $\hat{X}$ are  read off directly from the value of the quantity in this gauge, $\hat{X}\sim \check{X}$. Because of (\ref{massive Einstein equation}) the condition 
\beq \frac{[\hbar c]}{\alpha \xi ^2 |\check{\phi}|^2} \doteq \frac{16 \pi G}{[c^4]}, \qquad \mbox{with Newton constant $G$,} \label{transition to Newton constant}
\eeq 
 fixes the global constant of scalar field gauge  naturally.  Then $\check{\varphi} \neq 0$, if the scalar field itself is not already trivial, i.e., constant in Riemann gauge. IWG shows different features from Riemannian geometry! 
\end{itemize}
 In the first case  (Riemann gauge), Einstein gravity arises as  special case for a constant scalar field, in the second one  (scalar field gauge) for a vanishing scale connection. In both case  coherence is established if    $\frac{[\hbar c]}{2 \alpha \xi ^2 |\phi|^2}=\frac{8 \pi G}{[c^4]} $.

\subsection*{3 \quad Extension to the electroweak sector}
\subsubsection*{3.1 \quad Standard model fields  in Weyl geometry}
The Lagrangians  of the present standard theory of elementary particles are scale invariant over Minkowski spacetime. As far as quasi-classical fields are concerned, i.e. before ``second'' quantization,  they can be generalized straightforwardly  to  scale covariant ``curved metrics''. In particular  they can be imported to  Weyl geometry and combined with WOD gravity without major { conceptual} problems.\footnote{\cite{HungCheng:1988,Drechsler/Tann,Drechsler:Higgs,Nishino/Rajpoot:2004}    and \cite[chaps. 4, 8.1, 8.4]{Blagojevic:Gravitation}.}

Some  technical work has to be done mathematically. First of all,  the {\em Dirac operator} has to be imported to the underlying ``curved'' metric. In the Riemannian case it is well known how this is done \cite{Weyl:1929Dirac,Drechsler/Mayer:1977} \cite[chap. 19]{Frankel:Geometry}. The construction  can  be adapted to the Weyl geometric case (see below). As a mathematical condition, we have to assume both space and time orientability of $M$. Moroever, 
$M$  has to be a {\em spin}  manifold, i.e. it has to admit a principal $SL(2,\C)$ fibre bundle  globally (e.g. $H_2(M,\Z_2)=0$),  otherwise Dirac operators exist only locally \cite[515ff.]{Frankel:Geometry}. If $M$ is spin,  we can reduce  the structure group of the tangent bundle $TM$ to  $SL(2,\C)$. 
For any representation $\rho$ of  $SL(2,\C)$ (e.g. the Dirac spinor representation $\rho_D : SL(2,\C) \rightarrow GL_4(\C) $) the associated bundle  can be constructed, the sections of these serve as spinorial ``wave functions'' $\psi$. (Local) trivializations of the spinor bundle go in hand (are associated) to those  of the tangent bundle, arising from a specification of local orthonormal tetrads. 

The Levi-Civita connection $\Gamma$ can be reduced to $SO(1,3)$; let us call it $\omega $. Then it has  values  in $so(1,3)$, given by coefficients   $(\omega^{i} _{ j})$ with regard to vector fields $u=u^je_j$ developed  with reference to orthonormal tetrads $e_j$ and their duals $e^j$ ($0\leq j \leq 3$).\footnote{Latin indices indicate coefficients with regard to the orthotetrads, Greek ones with regard to coordinate derivatives. Partial derivation with regard to the vector field defined by $e_j$ will be denoted by $\partial_j:= e_j$, the Minkowski metric (with regard to tetrad coefficients) by $\eta =(\eta _{ij})$.}

For  $M$   spin, $\omega $  can be lifted coherently to a spinor connection $\tilde{\Gamma }$ with values in the Lie algebra of the universal covering $SL(2,\C)$ of $SO(1,3)$.  With (generalized)  Dirac matrices $\gamma ^i$  (in case of $\rho =\rho _D $ the ordinary Dirac matrices), $\gamma _i = \eta _{ij}\gamma ^j$ and $\gamma ^{\mu}=e^{\mu}_j \gamma ^j$ etc., the Lie algebra of $\rho(SL(2,\C))$ has generators  $ \frac{\sqrt{-1}}{8}\gamma ^i \gamma ^j $ \cite[(19.55)]{Frankel:Geometry}.

The covariant derivative of $\psi$ with respect to the spinor connection $\tilde{\Gamma }$  becomes (still in the Riemannian context)
 \beq (\nabla_{\tilde{\Gamma }}\psi )_l = \partial_{l}\psi + \tilde{\Gamma}_l \psi = \partial_{l}\psi + \frac{i}{8} \omega ^{jk}_{l} \gamma_j \gamma _k  \psi \; ,
 \eeq 
 and the purely gravitational Dirac operator (with vanishing scale and internal connection) is
  \beq \Dirac \psi_ := i \hbar \gamma ^l (\nabla_{\tilde{\Gamma }}\psi )_l  \, .
 \eeq

Secondly  the {\em structure group} of the theory {\em has to be extended} in order to account for the  ``internal'' dynamical degrees of freedom. The electroweak (ew) interaction is characterized by  dynamical symmetries in  $G_{ew}=SU(2) \times U(1)_Y$, constituted by the {\em weak isospin} group  $SU(2)$ and the hypercharge group $U(1)_Y$. Chromodynamics (cd) presupposes  $G_{cd}= SU(3)$ as structure group of ``local'' physical automorphisms. 

Mathematically, the extension proceeds by  introducing   trivial principal fibre bundles with regard to the structure groups $G_{ew}$ and $G_{cd}$, and by 
 tensoring the Dirac spinor bundles with  the appropriate representation spaces  of the dynamical groups.\footnote{More precisely, chirality of the weak interaction presupposes a  chiral decomposition of Dirac spinors: $\psi_L= \frac{1}{2}(1-\gamma _5) \psi , \,   \psi_R= \frac{1}{2}(1 +\gamma _5) \psi  \, , $ 
$\gamma _5=  \left( \begin{array}{c c}
 &  1 \\
 1 &  \\
  \end{array}\right)  = i \, \gamma _0\gamma ^1 \gamma ^2\gamma ^3$. The tensorial coupling of internal dynamics to the ``external'' degrees of freedom (encoded by the spinor spaces) differentiates between the two chiral components. See below (\ref{L_Psi}).}
 For $G_{cd}$ the ``colour'' representation space is isomorphic to $\C^3$, whereas for  $G_{ew}$ the representation spaces are the  weak isospin spaces well known from the standard model of elementary particle physics:
 \begin{itemize}
\item  Isospin $I=\frac{1}{2} $, representation spaces isomorphic to $\C^2$ for the chirally ``left''-handed spinors $\psi_L$ of three generations with basis respectively $(\nu_g,e_g)$ for the leptons ($Y=-\frac{1}{2} $) and $(u_g,d_g)$ for quark flavors ($Y=\frac{1}{6} $), $g=1,2,3$.
\item Isospin $I=0$, representation spaces isomorphic $\C$ for the chirally ``right''-handed spinors $\psi_R$, basis respectively $(e^R_g)$, perhaps also $(\nu^R_{g})$, and $(u^R_g),\, (d^R_g)$, with properly adapted hypercharge ($Y=-1$ for $(e^R_g)$ etc.).
\end{itemize}

The sections in the tensorized bundles (the ``full'' wave functions) may be denoted by $\Psi=\psi \otimes \psi_{int}$ with $\psi$ the Dirac spinor contribution, $\psi_{int}$ the representation space of the internal dynamics. The values of $ \psi_{int}$ lie in representation spaces of $ G_{ew}\times G_{cd}$. 

In the following we restrict our considerations to the electroweak sector. It is the one which  links most directly to gravitation. Chromodynamics  can, in principle, be appended like in the flat (Minkowski) case, but subtleties like mixing of downlike quark states and of neutrinos, CP-violating phase etc. have to be taken into account.

The  electroweak  potential decomposes as $W+B$ where 
\begin{itemize}
\item  $W$ is a connection with values in $su_2$, $W=(W_{\mu})$, curvature $F_W=dW+ \sqrt{2}g [W,W]=
(W_{\mu \nu })$ globally in an $ad\, SU_2$-bundle,\footnote{With generators $\tau _{\alpha }= \frac{i}{\sqrt{2}}\sigma _{\alpha } $ of $su_2$, $(\alpha =1,2,3)$, $W_{\mu } = A^{\alpha }_{\mu }\tau _{\alpha } $ $W= A^{\alpha }_{\mu }\tau _{\alpha }dx ^{\mu }$. $F_W=W_{\mu \nu }dx^{\mu }dx^{\nu } =  A^{\alpha }_{\mu \nu }\tau _{\alpha }dx ^{\mu }dx^{\nu }$ with $ A^{\alpha }_{\mu \nu }= \partial _{\mu }A^{\alpha }_{\nu } -  \partial _{\nu }A^{\alpha }_{\mu } $. This presupposes implicitly a gauge fixing (mathematically a standard trivialization) of the ew principal bundle and its isospinor adjoints such that the charge eigenstates lie ``up'' $(1,0)$ and ``down'' $(0,1)$.}
\item and $B$  a connection with values in $u_1 $, $B=(i\, B_{\mu})$, curvature $F_B= dB = (i \, B_{\mu \nu })$.\footnote{$B_{\mu \nu }= \partial _{\mu }B_{\nu }-  \partial _{\nu }B_{\mu }$}
\end{itemize}
The dynamical covariant derivative (ew sector only) of an isospinor field $\psi_{I,Y}$ with representation characteristics (``charges'') isospin $I$ and hypercharge $Y$ is 
\beq (D_{ew}\, \psi_{I,Y})_{\mu} =  ( \partial_{\mu} + I g \, W_{\mu}  + Yg' \, B_{\mu} )\psi_{I,Y} \label{D_ew}
\eeq
with coupling coefficents $g, g'$ for the weak isospin and hypercharge 
interactions.\footnote{In the textbooks often the traditional value for hypercharge $\tilde{Y}=2Y$ is used, in order to safeguard the `historical' Gell-Mann-Nishijima formular $q= I_3 + \frac{\tilde{Y}}{2}$ for electric charge $q$ ($I_3$ 3-component of isospin). The hypercharge coupling factor in (\ref{D_ew}) then becomes $\tilde{Y}\frac{g'}{2}$. Obviously in the convention chosen here,  $q=I_3 + Y$.}

Finally, the whole  construction can be adapted to the {\em Weyl geometric} case   \cite{Drechsler:1991,Drechsler:Higgs,Blagojevic:Gravitation}.  Obviously the orthonormal tetrads are of Weyl weight $w(e_j)= -1$, $w(e^i)=1$. The tetrad related metrical quantities remain unaffected by rescaling,  $w(\eta )= w(\omega )=0$. Also the tetrad related Dirac matrices $\gamma _j, \gamma ^k$ have Weyl weight $0$, while $w(\gamma ^{\mu})=-1$. Spinors are scale transformed by weight $w(\psi)=-\frac{3}{2}$ in order to achieve a scale invariant Lagrangian. 
 Of course also the weights of the internal group connections and their spinors remain unaffected by rescaling,  $w(W)=w(B)=0$ etc.  \cite[92ff]{Blagojevic:Gravitation}. 

The Weyl geometric scale covariant  derivative $ D_{\tilde{\Gamma }}$, respectively the Dirac operator $\Dirac$, of Dirac spinors  becomes
\beqa D_{\tilde{\Gamma }} \psi &=& \nabla_{\tilde{\Gamma }}\Psi - \frac{3}{2} \varphi \otimes \Psi \\
 \Dirac \Psi &=& i \hbar \gamma ^l ( D_{\tilde{\Gamma}} \Psi )_l   \, . \label{Dtilde}
\eeqa
The  {\em ew-dynamical scale covariant  derivative}   $\tilde{D}$ is
\beq \tilde{D}\Psi= \tilde{D}( \psi \otimes\psi_{I,Y}) = D_{\tilde{\Gamma }}\psi \otimes  \psi_{I,Y} + \psi \otimes D_{ew}\psi_{I,Y} \; ; \label{dynamical scale covariant derivative}
\eeq 
sometimes it is written as 
\[ (\tilde{D}\Psi )_{\mu}= (\partial_{\mu} +(\nabla_{\tilde{\Gamma}})_{\mu} - \frac{3}{2}\varphi_{\mu} + I g \, W_{\mu}  + Yg' \, B_{\mu} )\Psi_{I,Y} \; ,
\]
 \cite{Drechsler:1991,Nishino/Rajpoot:2004}. 

\subsubsection*{3.2 \quad Extending the scalar field to the ew sector,  Lagrangians} 
It seems natural to extend also the scalar field $\phi$ of Weyl-Omote-Dirac gravity   by a weak isospin  tensor factor. The tensorized scalar field will be denoted by  $\Phi$. If its properties lead to empirically sound conclusions it represents  a {\em gravitationally coupled electroweak vacuum structure}.
Several authors have noticed   that  for weak isospin $I=\frac{1}{2} $ and hypercharge $Y=\frac{1}{2} $,   $\Phi$ gets formal properties  analogous to the Higgs field of the standard model \cite{HungCheng:1988,Drechsler/Tann,Drechsler:Higgs}. We follow these authors and assume  the extended scalar field  with values in the $(I=\frac{1}{2}, Y=\frac{1}{2} ) $ representation space of $G_{ew}$, isomorphic to $\C^2$,
\beq \Phi (x) = (\phi_1(x), \phi_2(x)) \, , \label{Phi}
\eeq 
where the basis is chosen such that $\phi_1$ and $\phi_2$ characterize the electrically charged, respectively the neutral state ($q=I_3+Y$).

Analogous to (\ref{dynamical scale covariant derivative}) its  ew-dynamical scale covariant derivative   is 
\[  \left(\partial_{\mu }-\varphi_{\mu} + \frac{1}{2}g W_{\mu} +  \frac{1}{2}{g}' B_{\mu}) \right)\Phi  \,  . 
\] 
To better acccount for th common factor $\xi^2$  for the extended Lagrangian like in (\ref{L_phi}) or (\ref{L_G_phi}),  we shall use 
 the  notation  
\beqarr   \tilde{D}_{\mu}\Phi := (\partial_{\mu }-\varphi_{\mu} + \frac{1}{2}\tilde{g} W_{\mu} +  \frac{1}{2}\tilde{g}' B_{\mu}) \Phi  \, .   \label{ew dynamical derivative Phi}\\
\mbox{with} \quad  \tilde{g} = \xi^{-1}g, \quad    \tilde{g}'  = \xi^{-1}g' \, . \nonumber
\eeqarr
 Then the  Lagrangian of the  extended scalar field $\Phi$ has a form like in (\ref{L_phi}),
\beq  \mathcal{L}_{\Phi} =\  \frac{\xi^2}{2}\tilde{D}_{\nu}\Phi^{\ast} \tilde{D}^{\nu}\Phi \sqrt{|det \, g|}\, , \label{L_Phi}
\eeq  
but   now it contains a coupling between scalar field and ew gauge fields, due to the ew scale covariant derivative (\ref{ew dynamical derivative Phi}). With the exception of the Weylian scale connection term $\varphi_{\mu}$ and up to the factor $\xi^2$ (which cancels with the $\xi^{-1}$ in the couplings for terms quadratic in electroweak potentials), the  Lagrangian (\ref{L_Phi})  comes  down essentially to that of the Higgs field in the standard model.

For  the  {\em conjugate} (respectively the {\em adjoint}) of {\em spinors}\footnote{Complex conjugation is denoted by $\psi \mapsto \overline{\psi}$ and transposition $\psi \mapsto ^t\hspace{-0.3em}\psi$  (in both tensorial factors separately). }   the usual convention   
\beq \Psi^{\ast}= \, ^t\hspace{-0.1em} \overline{\Psi}  \gamma ^0 \, \eeq
is applied. 
Using abbreviated notation  $\Psi = \Psi^{(f,g,t)}$ with indexes   $f=l,q$  for the family  (lepton or quark),  $g=1,2,3$ for the generation, and  $t=u,d$ for the type (``uplike'', ``downlike''  weak isospin $I_3$ eigenstates),\footnote{Like usual, electrons are represented as  ``downlike'', neutrinos as ``uplike'', i.e., $\Psi_L^{l,1}= \left( \begin{array}{c} \nu_e \\e  \end{array}\right)$ etc. As an
 ad-hoc construction, neutrino masses are dealt with like the quark up states, i.e., right handed neutrinos are  assumed. } 
the fermionic Lagrangian $\mathcal{L}_{\Psi}= {L}_{\Psi }\sqrt{|det \, g|}$   is given by\footnote{Yukawa Lagrange terms are essentially the same as in \cite{Nishino/Rajpoot:2004}, with conjugate terms  added (reality of Lagrangian) like in \cite{Drechsler:Higgs}. Compare  the slightly different Lagrangian in \cite{Meissner/Nicolai}. \label{fn Yukawa terms}}
\beqarr {L}_{\Psi }&=& \frac{i}{2} [\hbar c] \sum_{f,g}  \sum_{J=L,R} \left( \Psi_J ^{\ast}\, \gamma ^{\mu}\tilde{D}_{\mu}\Psi_J  - (\tilde{D}_{\mu }\Psi _J)^{\ast}\, \gamma ^{\mu}\Psi_J  \right)
\\
& & \qquad    -  \epsilon _{sig}\sum_{f,g}  \left(  y( \tilde{\Psi} ^{\ast}_L  \Phi \tilde{\Psi }_{R} + \tilde{\Psi} ^{\ast}_{R} \Phi^{\ast}\,  \tilde{\Psi }_L ) + y' ( \tilde{\Psi} ^{\ast}_L  \tilde{\Phi }\tilde{\Psi }_{R} + \tilde{\Psi} ^{\ast}_{R} \tilde{\Phi}^{\ast}\,  \tilde{\Psi }_L ) \right)   , \nonumber \\
& & \mbox{where}  \quad   \tilde{\Phi} = i \sigma _2 \overline{\Phi}    \label{L_Psi} , 
 \quad \mbox{in 'unitary' gauge'} \quad \tilde{\Phi}   \doteq       \left(  \begin{array} { c  }  \overline{\phi}_o  \\0 \end{array} \right) = \left(  \begin{array} { c  }  \phi_o  \\0 \end{array} \right)  \, . \nonumber   \eeqarr

 The covariant derivative $\tilde{D}$ of spinors is  that of (\ref{Dtilde}).  ${\Psi }_R$  runs through up and down states separately.  For the quark family,  $\tilde{\Psi}$ denotes  down types  transformed to the mass eigenstates by the inverse of the (unitary) Cabbibo-Kobayashi-Maskawa (CKM) matrix, uplike types remain unchanged. For the lepton family, $\tilde{\Psi}$ denotes 
Maki-Nakagawa-Sakate  (MNS) matrix transforms of the uplike types (neutrino mixing), while the downlike types are untransformed.  $y$ and $y'$ are Yukawa coupling coefficients for down and up states respectively,  for each set of $(f,g,t)$ one.\footnote{Counting of parameters gives: number of  Yukawa parameters $|f|\cdot |g| \cdot |t|=2\cdot 3 \cdot 2=12$, 3 each for CKM and NNS matrices, 2 for CP-violating phase and QCD vacuum angle, 
2+1 coupling coefficients for ew and cd interactions,    two essential  parameters for gravity ($\xi, \lambda _4$) (perhaps three, depending on whether  $\alpha $ is considered as accidental or not),   no separate Higgs coupling coefficients (see below); sum 25 (without $\alpha $). It is likely that the representation of neutrino masses is redundant.} 
It may be appropriate to concieve of the different generations $g$ as excitation modes of spinor fields of given fermionic character $f$ and type $t$.
 $\Psi _L, \Psi _R$ are ``left'' and ``right''-handed editions of $\Psi $, 
\[  \psi_L= \frac{1}{2}(1-\gamma _5) \psi , \qquad   \psi_R= \frac{1}{2}(1 +\gamma _5) \psi  \, , \] 
$\gamma _5=  \left( \begin{array}{c c}
 &  1 \\
 1 &  \\
  \end{array}\right)  = i \, \gamma _0\gamma ^1 \gamma ^2\gamma ^3$.\footnote{Here, of course, $1=\left( \begin{array}{cc}
 1&   \\
  &  1\\
  \end{array}\right)$.}

The Yang-Mills action  for the ew boson field  $(W,B)$ (similar for the chromodynamic field) is as usual
\beqarr   \mathcal{L}_{ew} &=&  \frac{1}{4}[\hbar c] \left(  tr(W_{\mu \nu }W^{\mu \nu })  + (i\,B_{\mu \nu })(i\,B^{\mu \nu })\right) \sqrt{|det, \, g|}  \\
&=& - \frac{1}{4} [\hbar c] \left(  A^{\alpha }_{\mu \nu } A^{\mu \nu }_{\alpha } + B_{\mu \nu }B^{\mu \nu } \right)  \sqrt{|det, \, g|} \nonumber
\eeqarr 
Interactions with the fermionic fields are built into the ew gauge components of  $\tilde{D}_{\mu}$ of (\ref{L_Psi}).

In a full account, strong interactions have to be included in $\tilde{D} \Psi$ by covariant derivative components with values  in $\mathfrak g_{cd}=Lie\, (G_{cd})= su(3)$ and strong charges as coupling coefficients.  Recent high precision lattice calculations show that about 99 percent of total energy/mass of the observable bound states of mesons and baryons is due to the dynamics of QCD    \cite{Fodor:mass,Wilczek:mass}.
In regions of high matter density, their influence on the right hand side of the Einstein equation will be considerable.

Clearly, the ew covariant derivative terms of (\ref{L_Phi}) lead to formal mass terms for the ew bosons, 
\beq \frac{1}{4} \xi^2 \tilde{g}^2 |\Phi|^2 W_{\mu}W^{\mu} \qquad \mbox{and} \qquad   \frac{1}{4} \xi^2 \tilde{g}'^2 |\Phi|^2 B_{\mu}B^{\mu} \, . \label{ew boson masses 0}
\eeq  
Assuming  that a  change of basis (Glashow-Weinberg `rotation' about the {\em weak mixing angle} $\Theta $)   in the Liealgebra $\mathfrak{g}_{ew}= Lie(G_{ew})$  transforms the generators $W_0, W_1,W_2$  of ${su}(2)$ and $B$ of ${u}(1)$  into physical states $W^{\pm}, Z^o$, the mass terms of the latter become 
\beq m_W^2 = \frac{g^2}{4}|\Phi_o|^2\, , \qquad m_Z^2 = \frac{g^2}{4 \cos{\Theta }^2}|\Phi_o|^2  \; ,  \label{ew boson masses}  \eeq
with $g= \xi \tilde{g}, g'=\xi\tilde{g}'$ and $\cos \theta = {g} \,(g^2 + g'^2)^{-\frac{1}{2} } $
 like in special relativistic field theory. This presupposes that 
the scalar field $\Phi$ acquires a ground state $|\Phi_o|$ like the one of the quartic ``Mexican hat''  potential in the known approach. In the next section we shall see that such a ground state  arises naturally from the coupling between the scalar field and Weyl geometric gravity (WOD).

Varying with regard to $\Psi^{\ast}$ and $\Psi$ leads to the Dirac equation and its adjoint, varying with regard to the ew potentials leads to the Yang-Mills equations. Variation with respect to $\Phi^{\ast}$ gives the scalar field equation.    Because of the Yukawa-like contribution in (\ref{L_Psi}),   $\Phi$      couples to  fermionic matter fields. 

\subsubsection*{3.3 \quad Higgs potential condition, scalar field and Weyl gauge}
The total 
 {scalar field action} for $\Phi$  contains  quartic and  quadratic terms from the modified Hilbert term in (\ref{L_G_phi}) and the quartic potential (\ref{L_V4})
\beqarr \mathcal{L}_{\Phi, tot} &=& \frac{\xi^2}{[\hbar c]}  \left(  \frac{1}{2}\tilde{D}_{\nu}\Phi^{\ast} \tilde{D}^{\nu}\Phi  -V(\Phi)   \right) \sqrt{|det \, g|}  \label{L_Phi_tot} \\
V(\Phi) &=& \epsilon _{sig} \lambda_4 |\Phi|^4 + \alpha R |\Phi |^2 \; \ \label{potential Phi}
\eeqarr
With $sig\,(g) = (1,3)$ and $ \epsilon _{sig}= +1$,  the scalar curvature   $R$ is $< 0$ for reasonable { cosmological models}, more precisely for all Robertson-Walker models with  ``expansion'', i.e., warp $a'>0$,   and spatial leaves of curvature  $\geq 0$  or moderately negative (permitted  amount depending on warp). 
\begin{itemize}
\item   For $\alpha >0$ and $ \lambda_4 >0 $ and $sig = (1,3)$, the scalar field of WOD gravity, extended to the ew sector,  has a natural 
``Mexican hat''  potential given by gravitational, respectively cosmological coefficients:  $V$  is biquadratic, $V(x)= a_4 x^4 + a_2 x^2$, with  sign combination ($a_4 > 0, a_2 <0$), which assures stability of fluctuations about local minima given by $V'(x_o)=0, x_0 \neq 0$, i.e., $x_0^2 = -\frac{a_2}{2a_4} $.   
\item The calculation of the last item can be done at each spacetime point ($x_o \mapsto |\phi_o(p)|$). Thus the  
potential acquires a minimum for $\phi_0$ with
\beq \frac{|\Phi_o (p)|^2}{[\hbar c]^2}  = -   \frac{\alpha }{2} \epsilon _{sig} R(p) \,\lambda_4 ^{-1} \sim |R(p)|\, \lambda_4 ^{-1}  \, . \label{potential condition}
\eeq 
This relation will be called the {\em potential condition} for the scalar field. It {\em holds in any scale gauge} and  does not presuppose a breaking of scale symmetry. This is an important difference of Weyl geometric gravity  to Riemannian gravity.\footnote{Early in the 1970s  ``breaking of scale invariance'' of a conformally invariant scalar coupled to gravity in a Brans-Dicke-like way has been studied by S. Deser. Deser introduced an explicit scale breaking term by  quadratic term of the potential $\mu^2\phi^2 $ with a constant  $\mu$ \cite[252]{Deser:1970}. }
\item  In  the preferred scale choice (\ref{scalar field gauge}), in particular, Weyl geometric scalar curvature $R$ is normed to a constant, if the potential condition is satisfied.
\end{itemize}

Equ.  (\ref{potential condition}) shows that $|\Phi_o|^2$ is proportional to $R$,  if $\Phi$ is subject to the potential condition $V(|\Phi_o|) \stackrel{!}{=} min$. In this way, the {\em ground state of the scalar field}, $\Phi_o$,  {\em adapts  to Weyl geometric scalar  curvature}.
 Starting from  the Lagrangian (\ref{L_Phi_tot}),  the potential condition (\ref{potential condition}) implies that the scale gauge in which scalar curvature $R$ is constant is the same as the one in which the norm of the scalar field ground state $\phi:= |\Phi_o|$ is constant. 
  In other words, the potential condition ensures {\em identity of  scalar field gauge and Weyl gauge}.

\begin{itemize}
\item In the sequel we shall {\em assume} that a ground state of the scalar field exists and is in the minimum of the potential, i.e. satisfies the {\em potential condition}  (\ref{potential condition}). 
\end{itemize}

The potential condition regulates the norm of the groundstate of the quasi-classical field $\Phi$ only.  Even if one  looks for a deeper, or more refined, physical understanding of it, the analogy to the standard model Higgs field is no longer helpful.  The scalar field $\phi$  and its extension to the electroweak sector, $\Phi$, carry features of an order parameter which may express a possible underlying quantum reality close to the Planck scale only. It does not seem adequate to quantize the scalar field $\Phi$ as a whole. 
Far away from Planck scale physics, and as long as we do not know more about this level of reality, we should be content with analyzing the consequences of the structural properties of $\Phi$ assuming (\ref{potential condition}) for its ground state $\Phi_o$  (its normalization in the $I=\frac{1}{2}, Y= \frac{1}{2}  $ representation of  $SU(2)$ is discussed at the beginning of the next section).\footnote{Fluctuations $\chi $  about the ground state $\Phi = \Phi_o + \chi $ can probably be dealt with as quasi-particles, comparable to phonons in solid state physics (section 4.3).}

This  result is of considerable import for Weyl geometric gravity. If Weyl-Omote-Dirac gravity holds,  (\ref{L_G_phi}), and couples to the electroweak sector with minimal assumptions expressed in the Lagrangians  (\ref{L_Phi}), (\ref{L_Psi}), the biquadratic potential for the scalar field (\ref{potential Phi}) follows. The ground state of the scalar field then determines  the scale invariant magnitudes of observables uniquely  (up to a global constant) (\ref{scale invariant magnitude})  and selects Weyl gauge as the one in which magnitudes are most directly read off. One even may, but need not, read this relation as a kind of  ``vindication'' of Weyl's claim that physical clocks  calibrate ``by adaptation'' to a local field constellation in his discussion with Einstein  about his early gauge theory   \cite{Weyl:46}, \cite[298ff.]{Weyl:RZM5}. 

Far away from dense mass concentrations, e.g. around  ordinary stars and galaxies,   empty space  is  usually modelled by the Schwarzschild solution of Einstein gravity. That remains a  good approximation in Weyl geometry; but long-range effects of WOD gravity, usually considered as ``cosmological'', have to be taken into account for a more precise determination of curvature. In particular in ``empty'' space regions of laboratory scale,  $R$ cannot be assumed to be 0 in Weyl geometry, but has to be assumed to be of a ``cosmological'' value $R \sim H_1^2$. Here  $H_1=H_o c^{-1}\sim 10^{-29} cm^{-1}$ is the observationally determined Hubble parameter at present time. The hypothesis of standard cosmology that the cosmological warp function (``space expansion'') is    miraculously frozen inside galaxies (``Einstein-Strauss vacuoles'' or other ad hoc modifications of Robertson-Walker solutions) looses any plausibility if considered from the Weyl geometric perspective  (see section 4.2).

\subsubsection*{3.4 \quad Can gravity do what the Higgs field is supposed to do?\footnote{A similar question was  asked  by \cite{Pawlowski:1990}, there referring  to a conformal theory of gravity. }}
By a point dependent   gauge transformation in $SU(2)$, a gauge transformation in the isospin part of the ew bundle,  the ground state $\Phi_o$ can be normalized to the form 
\beq \Phi_o \doteq  (0,\phi_o ) \, , \qquad   \phi_o=|\Phi_o| \in \R \, .
\eeq 
$|\phi_o|$ is expected to correspond to the   electroweak energy scale 
$v=\frac{2}{g_w}m_Wc^2 $.\footnote{$m_W$= mass of $W$ boson $\approx 80.42 \, GeV$, $g_w= \frac{g_e}{\sin{\Theta _w}} \approx 0.6295$, $g_e=e\frac{\sqrt{4\pi}}{\hbar c}\approx 0.302 $.} 
This gauge fixing must be the same as in (\ref{Phi}).\footnote{Otherwise $\Phi_o$ would carry non-vanishing electrical charge and internal interactions.} 
In special relativistic ew theory this choice  is called the {\em unitary gauge} of the scalar  field. 
The   isotropy group at each point is isomorphic to an $U(1) \subset G_{ew}$  and  operates on the first isospinor component only (for the representation with $I=\frac{1}{2}, Y = \frac{1}{2}$). It encodes the  electromagnetic gauge symmetry which allows for
 long-range electromagnetic (em) fields.\footnote{For the  `left' lepton representation with $I=\frac{1}{2}, Y=-\frac{1}{2}$ the electromagnetic group $U(1)_{em}$ operates on the second component only.}  

A classical derivation of the mass terms (\ref{ew boson masses})  is possible by developing (\ref{L_Phi}) along the operations in direction of  generators of the Lie algebra of $G_{ew}$ transversal to $U(1)_{em}$ \cite{Bleecker:Gauge}. In this respect, there are only minor  formal differences to the usual Higgs field approach.  ``Ontologically'' and, as we shall see, experimentally, the difference is  considerable. In our approach, the scalar field expresses a $G_{ew}$-extended part of the gravitational structure; 
mass terms of the bosons (\ref{ew boson masses 0}) arise from coupling to gravitation, as it should be. In the standard view the nature of the Higgs field is wide open to physical speculation and philosophical controversy 
\cite{Earman:symmetrybreaking,Lyre:Higgs,Smeenk:Higgs}. Even in  a cautious interpretation,  the least one can say is that the extended Weyl geometric scalar field $\Phi$ expresses a connecting link between the electroweak sector and gravity via $\phi = |\Phi|$ and (\ref{L_G_phi}).  
But can a Weyl geometric Brans-Dicke like scalar field play a role usually ascribed to the Higgs field? 

If so, the potential condition (\ref{potential condition}), the relation with the Newton constant (\ref{transition to Newton constant}) and   the equality of $|\phi_o|$ with the experimentally determined  electroweak energy scale 
\beq |\phi_o|= v \approx 246\, GeV\, , \qquad \frac{v}{\hbar c} \approx 1.3 \cdot 10^{16} cm^{-1}\, , \label{v}
\eeq  
give detailed information on the coupling factor $\xi$ in (\ref{L_G_phi}),
\[ \xi^2 = \frac{[\hbar c] [c^4]}{\alpha v^2 \, 16 \pi G}= \frac{1}{16\alpha \pi } \frac{E_{Pl}^2}{v^2}  \, . \]
Condition (\ref{v}) specifies an energy, respectively length, unit in scalar field gauge.  Together with the global convention (2) in section 2.1 it specifies cgs units in the Weyl geometric/gravity context uniquely.\footnote{Of course this is a specification on the level of theoretical principles. For metrological purposes a technical better controllable specification will  be chosen, e.g., the one proposed for the new SI base units: The second, $s$, is set by ``fixing the numerical value of the ground state hyperfine splitting frequency of the caesium 133 atom, at rest and at a temperature of 0 K, to be equal to exactly 9192631770'' if  expressed in $s^{-1}$ \cite{BIPM:2010}.} 

 For $\alpha \sim 1$,  $\xi $ turns out to be essentially the {\em hierarchy factor} between Planck and electroweak energy scales,
\beq \xi = \frac{1}{\sqrt{16 \alpha \pi}} \frac{E_{Pl}}{v} \sim 10^{16} \, . \label{hierarchy factor}
\eeq 
Looked at it the other way round, $v$ is determined by the hierarchy factor $\xi$ in the Lagrangian (and $E_{Pl}$).

Similarly,  (\ref{potential condition}) and (\ref{v}) allow to  determine the relative orders of magnitude of $\lambda_4$ and $R$.  For cosmological curvature $R \sim - H_1^2 \sim - 10^{-56}\, cm^{-2}$ the value of  $\lambda_4$ turns out to be very small, 
\beq  \lambda_4 \approx - \frac{\alpha }{2} \frac{(\hbar c)^2}{v^2} R \sim 10^{-88}\, . \label{lambda-4}
\eeq

Although at a first glance, this seems inplausible because of its extreme smallness,   a look at  (\ref{L_G_phi}) shows that such an impression depends on comparing  incomparables. Rather than comparing $\lambda $ with 1, the values of $R$ and $\lambda_4 |\phi_o|^2 $ have to be put side by side.  (\ref{potential condition}) shows that this comparison fares quite well,  $|R| \sim H_1 ^2 \sim  10^{-56}\, [cm^{-2}]$ (in cosmologically estimated values) and $\lambda_4 |\phi_o|^2 \sim 10^{-88 + 32}$ [in $cm^{-2}$]  by (\ref{v}).

The last few paragraphs pursued a line of  analysis. The   ``ontologically'' more appropriate  synthetis  reverses the direction  of the argument.  We start from coefficients $\xi \sim 10^{16}, \,\lambda_4 \sim 10^{-88} $, find from cosmological observations $|R| \sim H_1 ^2 \sim  10^{-56}\, [cm^{-2}]$ and conclude from (\ref{potential condition}) $\phi_o^2  \sim \hbar c\,  |R| \lambda_4^{-1} \sim (100\, GeV)^2$ etc.\footnote{$\lambda _4 $ determines (or is determined by) the squared proportion between the electroweak distance scale $ \hbar c\, v^{-1} $ (roughly the ``smallest'' directly experimentally  accessible distance) and  Hubble length $H_1^{-1}$  (roughly the    ``largest''  distance indirectly accessible to astrophysical observations).}

On the other hand,  for fluctuations $\chi $ about the ground state, $\phi = \phi_o+ \chi $, a classical mass term  of the $\phi$-field may be developed from the seemingly ``tachyonic''  mass-like factor  $\alpha R$  ($R<0$) of the quadratic term in (\ref{potential Phi}).  Like in the ordinary ``Higgs'' model, and by the same calculation, the mass factor  turns sign in such a development, due to the contribution of the quartic term,  $\frac{1}{2}m_\chi ^2 \, [c^4]= - [\hbar c]^2\,  2 \alpha R $. That  results in
\beq m_\chi c^2  =2 [\hbar c]  \sqrt{ -  \alpha R} \sim 10^{-34} \, eV \,  \qquad \mbox{for $R \sim H_1^2, \, \alpha \sim 1$.}
\eeq 
That is a ridiculously small value. In fact it is the smallest amount of energy    that might be considered meaningful  in the cosmos.\footnote{The Compton wave length of $ m_\chi$ is at the order of magnitude of the Hubble length.} 

Quantum corrections may raise this value considerably. A very rough heuristic first estimation of  $\Delta m^2   \sim \lambda  \Lambda ^2 $,  with $\lambda := \xi ^2 \lambda _4$ the full (classical) quartic coefficient and with an energy cutoff at the order of Planck energy  $\Lambda \sim 10^{28}\, eV$,  would  raise the self energy of scalar field quantum fluctuations to   
\[ \Delta m^2 \sim 10^{-56+2\cdot28}\, eV^2 \sim  1\, eV^2 \, .  \]
This is  still far outside the  theoretical and experimental limits accepted in  the standard model.  In this  view such small values are theoretically excluded  because
 a Higgs mass below $100\, GeV $ would lead to instabilities at high energies  in the  ``early stage'' of the universe \cite{Espinosa_ea:Higgs_mass}.  Experimental exclusion below $114\, GeV$ is inferred from accelerator experiments at CERN's large electron collider LEP \cite{Aleph_ea:Higgs_LEP}. 

These limits are set from inside the framework of the standard model. It seems, however, unlikely that collider experiments would be able to ``see'' extremely small mass values  of  scalar field excitations as above. The theoretical exclusion, on the other hand, depends on the assumption of the correctness of the evolutionary picture of the early  universe. Its reliabilty may have to be reconsidered  if modifications of the standard model become necessary. For the moment, however,  the question in the title of this subsection has to be answered by a qualified negation: Gravity cannot do without basic modifications of the standard model what the usual Higgs is expected to do, although in principle it may  well serve for the generation of mass terms of fermions and electroweak interaction bosons.

 \subsection*{4 \quad Resum\'e and discussion}
  \subsubsection*{4.1  Reduction of symmetry}
  
All this appears surprising, in  its closeness and contrast to the established perspective on ew symmetry breaking. Different authors have developed other views on this topic from the point of view of Weyl geometry. Already since the 1970s   scale covariance of the theory and ``breaking'' of scale symmetry of the fundamental fields played a central role in considerations of what eventually became to be  called the ``Higgs mechanism'' 
\cite{Englert/Gunzig:1975,Smolin:1979,HungCheng:1988}.
Cheng considered the Weyl geometric scalar field $\Phi$ as ``Higgs field'' and ``broke'' its  scale covariance by  setting its norm (considered to be the ``expectation value'' of a not yet existing quantum theory) to the electroweak scale $v$.  He did not   discuss  conditions and reasons for that move. Neither the special role of the scalar field gauge for determining scale invariant magnitudes nor the potential condition came into sight; so his ``symmetry breaking'' must have looked  arbitrary to readers not well acquainted with Weyl geometry.  But  he realized  that (\ref{L_G_phi}) ``may form a basis for unification of the gravitational interaction with the electrowak interaction''  \cite[2183]{HungCheng:1988}.

For many researchers with a background in  differential geometry  of  fibre bundles, the whole discussion on symmetry breaking in field physics remained dubious. A. Trautman gave a clear analysis that,  from this point of view, talking about ``breaking'' of a symmetry from $G$ to $G' \subset G$ ought to be reserved for cases where a {\em reduction of the structure group} in the fibre construction is possible. Then the gauge group can be reduced to a ``localized'' symmetry in $G'$ (values of the connection in $\mathfrak g '=Lie\, G'$). Geometrically that presupposes a condition on the holonomy group  $H$ of the connections appearing in a dynamical theory (an integrability condition for the respective fields), $H\subset G'$. A special subcase is  an {\em explicit breaking} of the symmetry by a Lagrangian with smaller symmetry \cite{Trautman:1970}. 

In   Trautman's sense,  ``ew symmetry breaking'' carries  both aspects in a seemingly paradoxical way: no breaking (no reduction) on the ew distance scale itself, but breaking, in the sense of group reduction, on distance scales much larger. On larger scales, the effective curvature of the ew connections is negligible exactly because of the massiveness of the ew bosons,  i.e., as  a {\em result} of the Higgs procedure.\footnote{Cf. footnote \ref{fn Riemann} and the text above it.} 
On the ew distance scale, however,  where the Higgs ``mechanism''  takes place, symmetry in the sense of Trautman is {\em not} broken.  To the contrary, it is even  {\em  important that it is not}. The gauge group remains $G=SU(2)\times U(1)_Y$ after introducing the  scalar field. 

Of course, the  symmetry of $\Phi$, in mathematical terminology the isotropy group $G_{\Phi}$ of $\Phi$, is smaller than $G$, here $G_{\Phi}\cong U(1)_{em}$. But the gauge group still has local values in $G$ and cannot be reduced, as long as the ew field strengths do not vanish. Clearly, this is not the case: {\em no ew field strengths no ew bosons}, let alone massive ones. Moreover, for the formalism of the ordinary ``Higgs mechanism'', the transition to unitary gauge of the scalar field, $\Phi \doteq (0,\phi_o)$, is essential. Being able to do so presupposes that the full gauge group is still applicable. From the geometrical (fibre bundle) point of view,  no reduction of the gauge symmetry can be diagnosed, but rather a ``different realization'' of it  \cite{Drechsler:Higgs}. The procedure comes down to  representing the gauge connection at  the local values of $\Phi$ (operating in each infinitesimal neighbourhood of  $x$ close to  $\Phi(x)$).

Our view is different from Drechsler's and Tann's and from the classical Higgs mechanism, although it  overlaps with both in certain aspects. The introduction of scale invariant observable magnitudes, including those for mass parameters, equ. (\ref{scale invariant magnitude}),    makes ``breaking'' of scale symmetry  unneccessary on the quasi-classical level. On the classical level the  induced   { preferred} scale gauge, equ. (\ref{scalar field gauge}), serves as a tool for simplified calculation only. Although it allows to read off (or to plug in) classical mass values directly, it does not express an  explicit or  spontaneous breaking assumed by Drechsler/Tann or in the usual Higgs mechanism. On the other hand, it would not be surprising  if also in the Weyl geometric setting, like in other scale invariant approaches, the implementation of scale symmetry runs onto an anomaly during perturbative quantization, i.e.  cannot be  consistently expressed on the quantum level. Whether  this has to be interpreted as an indicator of ``spontaneous symmetry breaking'' in the usual sense  seems doubtful, considered for itself. Far below  the Planck level (energetically),   scale symmetry is no longer dynamical anyhow (section 2.3). Independent of any substantive realistic interpretation, a  restriction to the preferred scale gauge (\ref{scalar field gauge}) would just mean  that quantization presupposes an appropriate scale fixing on the classical level. The question of possible ``dynamical symmetry breaking'' close to the Planck scale latter is a different question; it may  be that here  't Hooft's idea of naturalness turns out to be a fruitful guide \cite{Hooft:naturalness}.  

Recently several authors have started to reconsider  scale covariant approaches to ew symmetry breaking, sometimes conformal, sometimes explicitly Weyl geometric \cite{Nishino/Rajpoot:2004,Nishino/Rajpoot:2007,Foot:2006,Foot:2007,Meissner/Nicolai}\footnote{In the 1990s  other authors continued the conformal approach, among them \cite{Pawlowski/Raczka:1995,Pawlowski/Raczka:1995_II}}
Closest to our and Cheng's view is the approach of Nishino/Rajpoot \cite{Nishino/Rajpoot:2004,Nishino/Rajpoot:2007}. These authors, however,  stay closer to the ordinary Higgs mechanism and allow  a  mass possibly observable in LHC experiments.  They do so by introducing  a second (real) scalar field $\sigma $, in addition to $\Phi$, also of weight $w(\sigma )=-1$, and arrive at a more flexible Lagrangian, with a scale invariant term of form $\mu \sigma |^2 |\Phi|^2 |$. That leads to an adaptable mass term $\mu \sigma |^2 |$ for $\Phi $, far above the tiny value induced by the purely gravitational quadratic term considered here.\footnote{Nishino/Rajpoot  compare  their approach with others in \cite{Nishino/Rajpoot:2007comment}.}

 For the moment, i.e., as long as no definite experimental results on the Higgs boson mass are available,    the most  simple form of the  Weyl geometric Lagrangian for the  scalar field, like in (\ref{L_G_phi}), remains  a  possibility  for giving mass to the ew bosons. Outright dismissing it because of its surprising effects
 would be  unjustified and at least premature. One may even appreciate  this approach, because it avoids the vexing  problem of compensatory terms for the  self energy corrections to the Higgs boson mass. It gives, at least, a conceptually pleasing and, if we are lucky, also a physically convincing account of the acquirement of mass by coupling to gravity. 
Of course, tracing  a  Higgs mass   at the LHC  scale would    experimentally refute the approach given here. 

If, on the other hand,  the Higgs mass remains elusive to the LHC detectors, the simple Weyl geometric   approach to gravity discussed here may show a well prepared path out of the otherwise  expected paradoxies for the standard theory of elementary particles.  It shows  that we need not necessarily  take  refuge to models in ``higher dimensions'',  multiplying the number of ``Higgs fields'', or other highly underdetermined  escape  routes of theory. We can  just as well look for a better understanding of the relation between gravity and ew interactions.

   \subsubsection*{4.2 \quad  Consequences for long range effects}
   If the scalar field plays a material role in gravity, and if the potential condition regulates its ground state, the consequences for considerations on large astrophysical scales, galaxies and above, and even for cosmology will be considerable. Not so, however, on smaller scales like the solar system, as the Schwarzschild solution is a degenerate vacuum solution of the WOD Einstein equations, with  $\phi = const$ in Riemann gauge and vanishing energy momentum tensor. 
    
Homogeneous and isotropic cosmological models behave slightly  differently  in the Weyl geometric appraoch.  Their  geometry in a manifold $M\approx I\times S^{(3)}$, with $I \subset \R$ and $ S^{(3)}$ a threedimensional manifold, can be
characterized by a classical {\em Robertson Walker metric} with spacelike folia of constant sectional curvature  and metric $a(\tau)d\sigma^2_{\kappa} $, where $d\sigma^2_{\kappa}$ denotes the 3-dimensional metric of constant sectional curvature $\kappa$ and $a(\tau )$ a warp function (``expansion'') depending on the cosmological time parameter $\tau$:
\beqarr \tilde{g}: \quad   d\tilde{s}^2 &=& - d \tau^2 + a(\tau )^{2} d\sigma^2_{\kappa}  \label{warped metric} \\
 d\sigma^2_{\kappa} &=&  \frac{dr^2}{1-\kappa\, r^2} + r^2 ( d\Theta ^2 +  r^2 \sin^2\Theta \, d\phi^2 ) \nonumber
\eeqarr
Looking at it in the Weyl geometric paradigm means to consider $\tilde{g}$  as  Riemannian component  of an IWG given by $(M, [\tilde{g},0])$.

If we assume a scalar field $\tilde{\phi}$ obeying the potential condition, the scalar field gauge is different from Riemann gauge. In fact, it is equal to Weyl gauge, as we have noted in (\ref{potential condition}). 
In Riemann gauge $(\tilde{g},0)$ we have
\beq \tilde{R} \doteq \,  _g\hspace{-0.1em}\tilde{R} \doteq -6\left( \frac{a'^2}{a^2} + \frac{\kappa }{a^2} +\frac{a''}{a}     \right) \, .
\eeq
So roughly $|\tilde{R}| \approx -6 H^2$,  for $\kappa \approx 0, |\frac{a''}{a}| \ll |a'|$,  with $H = \frac{a'(\tau _o)}{a(\tau _o)} $ at ``present''.  

Of course, the scaling condition for matter under expansion has to be related to  scale invariant magnitudes, the ``observables'' of  section 2.3. Thus different models from those of standard cosmology have to be taken into account.\footnote{A first exploration  can be found in \cite{Scholz:FoP} (still without using the potential condition). Using the potential condition, the scalar field energy momentum tensor  ensures dynamical consistency  (equilibrium) for the static ``Weyl universes'' discussed in \cite{Scholz:FoP}, if a homogeneously distributed mass density $\mu$ is assumed, for parameters $\alpha =1, \varsigma =3.5, \, \Omega _m\approx 2.83$. This does not seem  particularly convincing from the empirical point of view. Surely, dynamical considerations have to be added. }
As an important consequence of Weyl geometry, redshift (cosmological and gravitational) is invariant under change of scale gauge \cite{Scholz:FoP}. 

In Weyl gauge,  a considerable contribution to cosmological redshift will be due to the time component $\check{\varphi}_o$ of the scale connection $\check{\varphi}$. The same holds  for scalar curvature $\check{R}$. There is no reason, nor even any plausibility,  to assume a vanishing scale connection near  galaxies. This is an important difference to the received view in which ``expansion'' of the spatial folia is usually thought to  reach 
standstill close to galaxies. The   Einstein-Strauss vacuoles  which are often referred to in this respect, have to be considered an ad-hoc construction designed to bring about some kind of coherence between the usage of the Scharzschild solution close to stars and the expanding  space model of cosmology.\footnote{For a critical discussion of Einstein-Strauss vacuoles and other attempts to relate globally expanding cosmological models with local geometry close to stars see \cite{Giulini:EinsteinStrauss}.}
\begin{itemize}
\item In the Weyl geometric approach,  seemingly cosmological effects represented in Riemann gauge by the warp function $a(\tau)$ in (\ref{warped metric}) and  interpreted in terms of  an expansion of space in the standard approach, turn out to be due to ordinary, though {\em very weak,  field theoretical properties}.   In scalar field gauge,  they are mainly expressed  by the scale connection. 
 
\item If this approach is realistic, the corresponding effects are not truely cosmological in nature.   They are just so weak that they  were  first observed over very long range, i.e. on cosmological scales (in particular Hubble redshift with Hubble parameter $H$).
\item  There is no sound reason to assume that these effects  are shielded away by some unknown ``mechanism'' close to galaxies. Rather we have to assume that they   {\em pervade empty space everywhere}.  In particular, the curvature contribution of the warp function to scalar curvature  $R$ has to be considered as field theoretic property of the gravitational vacuum.
\end{itemize}
Thus the gravitational background should also be ``felt'' even in vacua underlying interactions of elementary particle fields, like in (\ref{potential condition}).

\subsubsection*{4.3 \quad Remarks on the relation between gravity and QFT}  
Up to here, we have approached the interface between gravity and quantum field theory (QFT) only marginally, basically from the classical side. The question of how to integrate the two sister theories of the 20th century about  spacetime and matter more deeply has not yet been dealt with. In the framework of our approach, a  natural  next step would be to  investigate quantized  versions of the  quasi-classical (spinor and connection) fields. That might be done perturbatively and/or by adapting methods of axiomatic and algebraic quantum field theory  to curved spaces. In both methods (and others) a systematic quantization of the gravitational field itself has, up to now,  resisted all attempts to come to definite results.

In our context, it  may be worthwhile to look at a somehow moderate quantization of gravity. Why should we assume that it is necessary  to quantize the  metrical structure  as a whole on all levels of energy?  
Years ago, Flato, Simon, and  Ra\c{c}zka have started to formulate a generalization of the Wightman axioms to curved spaces, slightly different from the one used today in theories of QFT on ``curved spaces'' \cite{Wald:QFT,Baer/Fredenhagen}. They foresaw a quantization of  the scaling degree of the metric and, by this,  a weak form of backreaction of quantum matter on the metric, while leaving the underlying conformal structure unquantized \cite{Flato/Simon,Flato/Raczka}. To my knowledge this approach  was not pursued further.
In our approach one might consider the Weyl geometric structure to be determined by the environmentally decohered classical matter systems, while only the scalar field degree of freedom of fluctuations about the ground state is subjected to quantization.

In WOD gravity the scalar field $\Phi$ is   directly linked, via (\ref{scalar field gauge})  and (\ref{scale invariant magnitude}) to the scaling degree of freedom of the Weylian metric.  
{\em Quantization of the scalar field}  therefore comes   down  essentially  to {\em quantizing the scaling degree of freedom} of the Weylian metric. Quantum field theory of a scalar covariant field  is  technically  much better  accessible than quantization of the complete metric and should be tractable by perturbative methods, also on curved spaces. Moreover, the scalar field seems to represent a material aspect of the extended gravity 
structure. 

There is  much room for investigations between the electroweak energy level and the Planck scale. It may well be that a  moderate quantization, concentrating on the scaling degree of freedom of the metric, suffices to understand many of the effects in the  energy regime   accessible during the next decades to  laboratory devices (LHC etc.) and astrophysical observation (radio, X-ray  telescopes etc.).

\subsubsection*{Acknowledgements} This work owes much to the discussion in the group of  historians/philosophers  of science and physicists cooperating in the  interdisciplinary  center of Wuppertal University and the project {\em Epistemology of the LHC}. Special thanks go to  Dennis Lehmkuhl for his detailed comments on the first draft of this paper and to the anonymous referee for his  advices how to improve the readability of the article.

\footnotesize
 \bibliographystyle{apsr}
\bibliography{alitmathsci,alithist}

\end{document}